\documentclass[pra,twocolumn, secnumarabic,nobalancelastpage,amsmath,amssymb,nofootinbib]{revtex4}

\usepackage{chapterbib}
\usepackage{color}
\usepackage{graphics} 
\usepackage{graphicx} 
\usepackage{subfigure}
\usepackage{longtable} 
\usepackage{bm} 
\usepackage{epstopdf}
\newcommand{\beq}{\begin{equation}}
\newcommand{\eeq}{\end{equation}}
\newcommand{\bqa}{\begin{eqnarray}}
\newcommand{\eqa}{\end{eqnarray}}
\newcommand{\nn}{\nonumber}

\newcommand{\BQIC}{Berkeley Center for Quantum Information and Computation, Berkeley, California 94720 USA}
\newcommand{\DeptPhys}{Department of Physics, University of California, Berkeley, California 94720 USA}
\newcommand{\DeptChem}{Department of Chemistry, University of California, Berkeley, California 94720 USA}

\newcommand{\erf}[1]{Eq.~(\ref{#1})} 

\begin{document}

\title{What is the optimal way to prepare a Bell state using measurement and feedback?}

\author {Leigh Martin$^{1,2}$}
\email {Leigh@Berkeley.edu}
\author{Mahrud Sayrafi$^{1}$}
\author{K. Birgitta Whaley$^{1,3}$}

\affiliation{$^1$\BQIC}
\affiliation{$^2$\DeptPhys}
\affiliation{$^3$\DeptChem}

\date{\today}

\begin{abstract} 
Recent work has shown that use of quantum feedback can significantly enhance both the speed and success rate of measurement-based remote entanglement generation,
but it is generally unknown what feedback protocols are optimal for these tasks.  Here we consider two common measurements that are capable of projecting into pairwise entangled states, namely half- and full-parity measurements of two qubits,
and determine in each case a globally optimal protocol for generation of entanglement. 
For the half-parity measurement, 
we rederive a previously described protocol using more general methods and prove that it is
globally optimal for several figures of merit, including maximal concurrence or fidelity, and minimal time to reach a specified concurrence or fidelity. 
For the full-parity measurement, we derive a protocol for rapid entanglement generation related to that of (C. Hill, J. Ralph, Phys. Rev. A 77, 014305 (2008)), 
and then map the dynamics of the concurrence of the state to the Bloch vector length of an effective qubit. This mapping allows us to prove several optimality results for feedback protocols with full-parity measurements. We further show that our full-parity protocol transfers entanglement optimally from one qubit to the other amongst all measurement-based schemes. The methods developed here will be useful for deriving feedback protocols and determining their optimality properties in many other quantum systems subject to measurement and unitary operations.
\end{abstract}

\maketitle

\section{Introduction}
\label{sec:introduction}

The non-local nature of quantum entanglement has played a major role in the theoretical development of quantum mechanics, and is anticipated to play an equally significant part in its practical applications.  
Remote entanglement is a necessary resource for many quantum technologies, such as quantum cryptography, distributed and blind \cite{broadbent2009universal} quantum computing, and general quantum networking. Such applications may take place over kilometer length scales, but intermediate ranges are also relevant. For example, fault-tolerant quantum computing may require hundreds of millions of qubits, and the resulting device could be extremely large in size \cite{Fowler2012surface}. To construct and characterize such a system, it is likely to be built in a modular way, much like a modern supercomputer \cite{Devoret2013superconducting, Benjamin2013}. 
Efficient distribution of entanglement between spatially separate parts of such a system will then be an architectural requirement that presents a substantial engineering challenge. 

Of the existing experimental schemes for entanglement generation between quantum objects that are too far apart to directly interact, many are based on measurement \cite{Motzoi2015, Roch2014, Hofmann2012, Bernien2013, Vcernotik2015measurement, Roy2015remote, Silveri2016theory} and dissipation \cite{Motzoi2015back}, rather than relying on quantum gates. Such schemes are more robust to loss, which tends to increase with separation distance. By combining signals from two separate objects and measuring the combined signal, these schemes can project a system into an entangled state while only being sensitive to loss from one pass through the connecting channel. Although these methods are becoming ubiquitous in experimental quantum science, little is known about the optimal way to use them. Furthermore, deterministic as opposed to probabilistic generation of remote entanglement between stationary (\textit{i.e.} non-flying) qubits remains an outstanding experimental challenge. However, theoretical work has demonstrated that measurement-based feedback can be used to enhance the success rate and fidelity of measurement-based entanglement generation \cite{Sarovar2005, Hill2008, HPFPRA}. The fact that more entanglement can be created during the time over which measurement collapse occurs indicates that quantum feedback provides a fundamental advantage, but it is unknown what measurement-based feedback protocols are optimal for such tasks or which schemes are most efficient in principle.

Such optimization lies in the domain of quantum optimal control. However, to date only a limited number of measurement-based quantum feedback protocols have been established to be globally optimal, owing in part to the non-linear nature of the problems. A feedback scheme for qubit purification was discovered by Jacobs \cite{Jacobs2003} and its optimality properties were studied in a number of subsequent works \cite{Wiseman2006reconsidering, Wiseman2008, Teo2014, Li2013}. A protocol for rapid purification of a qudit was shown to exist in \cite{Combes:2006hx}, and constructed explicitly in \cite{Shabani2008locally}. Upper bounds on qudit and multiqubit purification speedups are known \cite{Combes2010rapid}, but proving global optimality for all situations remains an open challenge except for the qutrit case, which was solved in \cite{Shabani2008locally}. A globally optimal protocol for rotating a monitored qubit to a desired state was given in \cite{Sridharan2012optimal}, but the precise way in which it is optimal is somewhat unnatural and several basic questions remain open.

Regarding feedback for entanglement generation, some optimality results are known for linear quadratic Gaussian systems \cite{Serafini2010determination, Mancini2007optimal, Genoni2013optimal}. Feedback protocols for enhancing the rate of entanglement generation using half-parity and full-parity measurements were given in \cite{HPFPRA} and \cite{Hill2008} respectively, but without proof of optimality. Both protocols generate entanglement faster than is possible with measurement alone.
They are also of interest because they drive the system along a deterministic path in Hilbert space; feedback is chosen to exactly cancel the randomness introduced by measurement, allowing for deterministic entanglement generation. This is particularly significant for the half-parity measurement, which has only a 50\% success rate in the absence of feedback.

In this work, we prove that the half-parity measurement protocol given in \cite{HPFPRA} is globally optimal for several practically relevant tasks listed in the following section. For the full-parity measurement, we derive a protocol which is related to that given in \cite{Hill2008} but which is designed for initially pure states. Elaborating on the connection between rapid entanglement generation and rapid purification first discussed in \cite{Hill2008}, we derive a natural mapping between this two-qubit system and measurement of an effective qubit, in which the concurrence \cite{Wootters1998} exactly maps to this qubit's Bloch vector length. We then use this mapping to prove several optimality results for the proposed full-parity protocol.
We also show that our full-parity protocol 
is optimal among all measurement-based entanglement generation schemes that acquire single qubit information at a fixed rate. This result sets a useful standard by which the effectiveness of an entanglement generation protocol may be assessed. 

We begin in Sec. \ref{sec:Measurement} with an introduction to continuous measurements and list the optimality results to be proved. Section \ref{sec:Feedback} rederives the half-parity feedback protocol in a more general way and presents the full-parity feedback protocol. The main section of the paper is Sec. \ref{sec:GlobalOptProofs}, where we prove our optimality results. Sec. \ref{sec:Conclusion} discusses potential applications and a number of future directions. Details for implementation of the full-parity feedback protocol and its relation to that of Ref. \cite{Hill2008} are given in appendix \ref{sec:FPFDetails}, where we also elaborate on the connection between rapid purification and rapid entanglement generation.
 
\section{Non-local and Continuous Measurements}
\label{sec:Measurement}

The measurements we will consider are the full-parity measurement $X_F = \sigma_{z1}\sigma_{z2}/2$ which distinguishes odd- from even-parity states, and the half-parity measurement $X_H = (\sigma_{z1}+\sigma_{z2})/2$. The latter is so called because it does not distinguish the odd parity states $|01\rangle$ and $|10\rangle$, but does distinguish the even parity states $|00\rangle$ and $|11\rangle$. If two qubits are initialized in the separable state $(|0\rangle+|1\rangle)\otimes(|0\rangle+|1\rangle)$ (ignoring normalization), then a full-parity measurement collapses the state into either $|01\rangle+|10\rangle$ or $|00\rangle+|11\rangle$, both of which are entangled. In contrast, the half-parity measurement only creates entanglement with a 50\% success rate, because the separable outcomes $|00\rangle$ or $|11\rangle$ occur half the time. Despite this disadvantage, half-parity measurements are often easier to implement in practice. These operators have both been implemented as continuous measurements in superconducting qubits in references \cite{Riste2013} and \cite{Roch2014} respectively. Although the latter did not involve remote qubits, full-parity measurements can be implemented on remote qubits as well using the same experimental system \cite{Motzoi2015}. 

In the idealized limit of instantaneous projective measurements, the notion of applying feedback during a measurement is ill-defined and full-parity measurement alone therefore already creates perfect, deterministic entanglement with no time delay.
However realistic measurements occur over a finite duration as a continuous process which we describe in detail below. This provides the opportunity to apply feedback by performing additional operations to a system conditional on the measurement outcome up to that point in time. As these measurements are often resource-intensive, it is desirable to optimize the rate of entanglement generation.

In this work, we assume a controller has the ability to apply feedback only in the form of local unitary (LU) rotations on the measured qubits, so that measurement is the only source for entanglement generation. This restriction substantially simplifies experimental implementation and is essential when the qubits are remotely separated. 
The protocols for $X_F$ and $X_H$ are given in Sec. \ref{sec:Feedback}. We name these as $P_F$ and $P_H$ respectively. Each protocol involves applying proportional feedback, \textit{i.e.}, local unitary rotations with a rotation angle that is proportional to the most recent measurement outcome. In Sec. \ref{sec:GlobalOptProofs}, we prove the following global optimality properties:\\

\textit{Half-parity measurement} $X_H = (\sigma_{z1} + \sigma_{z2})/2$:

$P_H$ is globally optimal for the following tasks.
\begin{enumerate}
\item Max. concurrence goal: $P_H$ maximizes the expectation value of the entanglement (quantified using the concurrence $\mathcal{C}$ \cite{Wootters1998}) reached at a chosen stopping time $T$.
\item Max. fidelity goal: Same as above but for fidelity $\mathcal{F}$ \cite{Jozsa1994fidelity} with respect to any target state which is pure and maximally entangled.
\item Min. time goal for concurrence: $P_H$ minimizes the expected time to reach a desired concurrence $\mathcal{C}_\text{Threshold}$, so long as $\mathcal{C}_\text{Threshold}\leq 1/\sqrt{2}$.
\item Min. time goal for fidelity: Same as above but for a desired fidelity $\mathcal{F}_{\text{Threshold}}$, so long as $\mathcal{F}_{\text{Threshold}} \leq (1 + \sqrt{2})/\sqrt{8}$
\end{enumerate}

\textit{Full-parity measurement} $X_F = \sigma_{z1}\sigma_{z2}/2$:
\begin{enumerate}
\item Max. concurrence goal: $P_F$ is globally optimal.
\item Max. Fidelity goal: $P_F$ is globally optimal
\item Min. time goal ($\mathcal{C}$ or $\mathcal{F}$): Measuring without feedback is globally optimal.
\end{enumerate}

The min. time results for concurrence also apply to any monotonic function of $\mathcal{C}$, such as entanglement of formation. Finally, we show that $P_F$ is optimal for the max. concurrence goal among all protocols that acquire single qubit information at a fixed rate. A precise statement of this result is given with its proof in Sec. \ref{sec:GlobalOptProofs}.

  We now provide a brief introduction to continuous measurement. In both applications mentioned above, the dynamics of the monitored system can be modeled as continuous measurements. These are described by a stochastic master equation   ~\cite{Wiseman2009, JacobsSteck2006, jacobs2014quantum}:
\begin{align}
\label{eq:RhoMCont}
&\rho(t+dt) = \rho(t) + \mathcal{D}[M]\rho(t) dt + \mathcal{H}[M] \rho(t) \sqrt{\eta} dW(t) \nonumber \\
&M \equiv \sqrt{\frac{\Gamma}{2}}X,
\end{align}
where $X$ is either the full or half-parity measurement operator. Following convention, we have defined the Lindblad dissipator $\mathcal{D}[M] \equiv M \rho M^\dagger -1/2(M^\dagger M \rho + \rho M^\dagger M)$ and the measurement superoperator $\mathcal{H}[M]\rho \equiv M \rho + \rho M^\dagger - \langle M+M^\dagger \rangle \rho$. $\Gamma$ is the measurement rate, which sets the timescale for collapse onto an eigenstate of X. Lastly, $\eta$ is the measurement efficiency, which characterizes how much of the signal is collected and observed. $\eta$ may range from zero to one. Because measurement introduces randomness into the evolution of a wave function, the equations of motion must have a stochastic component, which in this case is given by $dW$, a Gaussian distributed zero mean random variable with variance $dt$. In this work, we treat this quantity according to the rules of Ito calculus, although the Stratonovic formalism may also be used \cite{Oksendal2003, Wiseman2009}.

  To study the dynamics of \erf{eq:RhoMCont}, one may use Monte Carlo simulation to sample $dW$ and generate possible realizations of the measurement outcome, usually termed quantum trajectories \cite{Wiseman2009}. In an actual experiment, $dW(t)$ is derived from the signal generated by the measurement apparatus, $dV_t$. The relation between this signal and $dW$ is \cite{jacobs2014quantum, JacobsSteck2006}
\begin{equation}
\label{eq:HomV}
dV_t=\langle X \rangle(t) dt + \frac{dW(t)}{\sqrt{2 \eta \Gamma}}.
\end{equation}
Feedback then consists of applying an additional Hamiltonian which depends on both the measurement record $dV(t')$ over the interval $0 \leq t' \leq t$, and the initial state. Equivalently, one can use \erf{eq:RhoMCont} and $dV(t')$ to estimate the current state and calculate feedback accordingly.

In what follows, we make the simplifying assumptions that $\eta=1$ and that the initial state is pure. In the absence of additional decoherence, the state remains pure at all times and one may then replace \erf{eq:RhoMCont} with a simpler stochastic Sch\"odinger equation
\begin{equation}
\label{eq:PsiMCont}
\psi(t+dt) = \Big[ -\frac{1}{2}(X-\langle X \rangle)^2 dt + (X-\langle X \rangle) dW(t) \Big]|\psi(t)\rangle,
\end{equation}
where we have set the measurement rate $\Gamma=2$ for simplicity, a convention that we will retain in what follows. \\

\section{Feedback Protocols for Entanglement Generation}
\label{sec:Feedback}

Our objective is to quantify how entanglement changes under arbitrary feedback protocols for binary systems, so that we can identify which are optimal. We characterize entanglement using the concurrence, which for pure states is defined as \cite{Wootters1998} 
\begin{equation}
\label{eq:Concurrence}
\mathcal{C} \equiv |\langle \psi^*| \sigma_y \otimes \sigma_y |\psi \rangle|
\end{equation}
with $\langle\psi^*|$ the complex conjugate of $\langle\psi|$, or equivalently the transpose of $|\psi\rangle$. Like all valid entanglement measures, $\mathcal{C}$ is invariant under local unitary rotations, so our allowed feedback operations leave it unchanged. For the time being, we take $\mathcal{C}$ to be our figure of merit. 

Since all bipartite pure states with the same concurrence are equivalent up to LU operations \cite{Nielsen2010}, it is possible to parameterize any pure state in terms of $\mathcal{C}$ and single qubit rotations. Such a parameterization is useful in this context because 
feedback can directly control the latter quantities. Therefore we can model feedback to directly set these qubit rotation parameters to desired values without specifying the Hamiltonian necessary to prepare the resulting state.
The Schmidt decomposition provides an explicit example of such a parameterization \cite{Nielsen2010}. By expressing the Schmidt coefficients in terms of the concurrence, we can write a general two-qubit state as
\begin{align}
\label{eq:StateParam}
\psi &(\mathcal{C}, \theta_1, \theta_2, \phi_1, \phi_2, \gamma_1, \gamma_2)  \\ \nn
&= U_1 \otimes U_2 \Bigg{[} \sqrt{\frac{1+\sqrt{1-\mathcal{C}^2}}{2}} |00\rangle - \sqrt{\frac{1-\sqrt{1-\mathcal{C}^2}}{2}} |11\rangle \Bigg{]} \\ \nn
&U_i \equiv \exp(-i \gamma_i \sigma_z/2) \exp(-i \sigma_y \theta_i/2) \exp(-i \phi_i \sigma_z/2), \\ \nn
\end{align}
where we have written $U_i$ in terms of the Euler angles $\{\phi_i, \theta_i, \gamma_i\}$. For convenience in subsequent calculations, we break the local unitaries into symmetric and antisymmetric rotations by defining $\theta \equiv (\theta_1 + \theta_2)/2$, $\Delta \theta \equiv (\theta_1 - \theta_2)/2$, and likewise for $\phi$ and $\gamma$. The final expression does not depend on $\Delta \phi$ because the state in brackets is invariant under antisymmetric rotations about $\sigma_z$. This separation of variables into an LU-invariant quantity depending on $\mathcal{C}$ and LU rotations depending on $\{\theta_i,\phi_i,\gamma_i\}$ motivates an analogy with parameterization of a qubit in terms of a Bloch vector $\vec{r}$ in spherical coordinates; $\mathcal{C}$ is analogous to $|\vec{r}|$, which is invariant under unitary operations, while $\{\theta_i,\phi_i,\gamma_i\}$ can be set arbitrarily using Hamiltonian feedback like $\theta$ and $\phi$. When considering the full-parity measurement in Sec. \ref{sec:GlobalOptProofs}, we will find that this analogy admits an explicit mapping in which $\mathcal{C}$ and $r$ obey the same equations of motion under measurement.

To study how entanglement changes under measurement, we substitute \erf{eq:StateParam} into \erf{eq:PsiMCont} and compute the concurrence of the resulting state $\psi(t+dt)$. It is not necessary to compute how the five angles parameterizing $\psi$ evolve, since we will model the controller to set them according to some feedback protocol. The computation is further simplified by the fact that $\sigma_z$ rotations commute with $X_F$ and $X_H$, so that the resulting equations of motion do not depend on $\gamma$ or $\Delta \gamma$. We first focus on the half-parity measurement $X_H$. As is detailed in appendix \ref{sec:CalcDetails}, application of Ito's lemma yields
\renewcommand{\arraystretch}{1.5}
\begin{align}
\label{eq:dCHPF2}
d\mathcal{C} = \left\{ \begin{array}{ll}
					2\mathcal{C}\sqrt{1-\mathcal{C}^2} u v~dW \\
					~~+ [(v^2-u^2)w-\mathcal{C}(v^2+u^2)]dt 	~~ &|~\mathcal{C}>0 \\
					|v^2-u^2|~dt	&|~\mathcal{C}=0
\end{array}
\right. 
\end{align}
where we have defined the control variables $u = \cos(\theta)$, $v = \cos(\Delta \theta)$ and $w=\cos(2\phi)$. 

In what follows, we implicitly assume that at any time, the controller can instantaneously set the angles $\theta, \Delta \theta$ and $\phi$ to any desired value. This is equivalent to assuming that the controller can implement any local unitary on both qubits with no time delay. For many systems such as superconducting qubits, single qubit rotations can be performed much faster than the measurement time scales, which makes this approximation appropriate. The assumption of zero time delay can be satisfied as long as the propagation delay between the qubits is small relative to the inverse measurement rate. However, this restriction can be relaxed for the full-parity protocol, as 
it will turn out that the protocol can be implemented by applying feedback on only one of the two qubits.
Thus, \erf{eq:dCHPF2} is an equation of motion for the concurrence of the state under arbitrary feedback protocols that may be specified by choosing a particular, set $u(t,\mathcal{C})$, $v(t,\mathcal{C})$ and $w(t,\mathcal{C})$. 

Although there is no reason \textit{a priori} that a locally optimal strategy is also globally optimal, it often turns out to be so in practice. This will be the case in the results that we prove here. The locally optimal protocol maximizes the expectation value of the concurrence at time $t+dt$. To find it, one simply chooses the values of $u$, $v$ and $w$ to maximize the $dt$ term of \erf{eq:dCHPF2}. This occurs for $\{u=0,v=1,w=1\}$ with all values of $\gamma$ and $\Delta \gamma$ allowed.\footnote{$\{u=1,v=0,w=-1\}$ is also a solution,  but since it is actually equivalent to the first if one makes the transformation $\gamma \rightarrow \gamma-\pi/2$, $\Delta\gamma \rightarrow \Delta \gamma + \pi/2$, we ignore it. Solutions corresponding to $u=-1$ or $v=-1$ are similarly redundant.} We henceforth refer to this protocol as $P_H$. The resulting equations of motion under this set of control parameters may be easily solved:
\begin{equation}
\label{eq:PH}
d\mathcal{C} = (1-\mathcal{C})dt ~~\implies \mathcal{C}(t) = 1-(1-\mathcal{C}(0))e^{-t}.
\end{equation}
The evolution of the state under this feedback protocol may be computed by substituting this solution and the above control values into \erf{eq:StateParam}. Setting $\mathcal{C}(0) = 0$ for simplicity, the state evolution under feedback is thus
\begin{align}
\label{eq:HPFPsiSol}
\psi(t) = \frac{1}{2}
 \begin{bmatrix}
e^{-i \gamma} \sqrt{e^{-t}}	\\
e^{-i \Delta\gamma} \sqrt{2-e^{-t}}	\\
e^{i \Delta \gamma} \sqrt{2-e^{-t}}	\\
e^{i \gamma} \sqrt{e^{-t}} 	\end{bmatrix}.
\end{align}
For $\gamma=\Delta \gamma = 0$, this state evolution exactly matches the solution to the feedback protocol given in \cite{HPFPRA}, indicating that they are the same feedback protocol. 

Equation (\ref{eq:HPFPsiSol}) can be more easily understood by writing down the states for $t=0$ and $t \rightarrow \infty$. Up to a global phase, these states are
\begin{align}
\psi(0) &= \frac{1}{2}\big(|0\rangle + e^{i\gamma_1} |1\rangle \big) \otimes \big( |0\rangle + e^{i \gamma_2}|1\rangle \big) \nn \\
\psi(\infty) &= \frac{1}{\sqrt{2}}\big( |01\rangle + e^{2i \Delta \gamma} |10\rangle \big).
\end{align}
Thus this protocol involves preparing both qubits in a separable state polarized along some axis of the equator of their respective Bloch spheres. Feedback deterministically projects the qubits into a Bell pair with the relative phase determined by the relative equatorial angle of the initial preparation. As shown in Ref. \cite{HPFPRA}, this protocol may be implemented with proportional feedback, \textit{i.e.}, by continuously applying local qubit rotations with an angle proportional to the measurement signal $dV$.

Before proving global optimality results for the above protocol, we first repeat the above analysis for the full-parity measurement. Following similar steps to those above but now for $X_F$, the equations of motion for the concurrence can be shown to be
\begin{align}
\label{eq:EOMsF}
d\mathcal{C}  = \left\{ \begin{array}{ll}
	 (1-\mathcal{C}^2)\big[(u^2-v^2)w~\mathcal{C}~dW \\ 
	~~~~ + (u^2-v^2)^2(1-w^2)dt/2\mathcal{C}\big]~~~	& |~\mathcal{C}\neq0 \\
					(u^2-v^2) dW	&  |~\mathcal{C}=0,
\end{array}
\right. 
\end{align}
A minor technicality in the derivation given in appendix \ref{sec:FPFDetails} has forced us to allow $\mathcal{C}$ to take negative on values in general. However since $\psi(\mathcal{C})$ only depends on $\mathcal{C}^2$, this fact presents no further difficulty. We shall simply interpret $|\mathcal{C}|$ as the concurrence instead of $\mathcal{C}$.

For the full-parity measurement, two distinct sets of locally optimal parameters emerge: $\{u=0,v=1,w = 0\}$ and $\{u=1,v=0,w=0\}$. The resulting state evolutions are equivalent up to a complex conjugation of $\psi$, so the underlying dynamics are therefore essentially identical. We may therefore only consider the protocol $P_F:~\{u=0,v=1,w = 0\}$ and ignore the other solution.
As for the half-parity case, the equations of motion for $\mathcal{C}(t)$ are deterministic and again yield an analytic expression with easy solution:
\begin{equation}
\label{eq:PF}
d\mathcal{C} =  \frac{1-\mathcal{C}^2}{2\mathcal{C}}dt ~~\implies \mathcal{C}(t) = \pm\sqrt{1-[1-\mathcal{C}(0)^2]e^{-t}}.
\end{equation}
Validity of this solution at $\mathcal{C}(0)=0$ is most easily established by deriving equations of motion for 
$\mathcal{C}(t)^2$ 
at $\mathcal{C}=0$ and showing that the solutions coincide. State evolution under this protocol is given by
\begin{equation}
\label{eq:FPFPsiSol}
\psi(t) = \frac{1}{\sqrt{8}}
 \begin{bmatrix}
e^{-i (\gamma+\pi/4)} \big(\sqrt{1-e^{-t/2}} - i \sqrt{1+e^{-t/2}} \big) \\
e^{-i (\Delta\gamma+\pi/4)} \big(\sqrt{1-e^{-t/2}} + i\sqrt{1+e^{-t/2}} \big) \\
e^{i (\Delta \gamma-\pi/4)} \big(\sqrt{1-e^{-t/2}} + i\sqrt{1+e^{-t/2}} \big)	\\
e^{i (\gamma-\pi/4)} \big( \sqrt{1-e^{-t/2}} - i\sqrt{1+e^{-t/2}} \big)	\end{bmatrix} .
\end{equation}
Again, taking early- and late-time limits of the solution gives some insight into the induced dynamics. Here
\begin{align}
\psi(0) &= \frac{1}{2}\big(|0\rangle + e^{i\gamma_1} |1\rangle \big) \otimes \big( |0\rangle - i e^{i\gamma_2}|1\rangle \big) \\
\psi(\infty) &= \frac{1}{2}\big( e^{-i\gamma} |00\rangle + e^{-i \Delta \gamma}|01\rangle + i e^{i\Delta \gamma} |10\rangle - i e^{i \gamma} |11\rangle \big), \nn
\end{align}
where as before we have dropped a global phase from the states. The optimal initial state is again to prepare both qubits in an equal superposition of $|0\rangle$ and $|1\rangle$. Somewhat counterintuitively, the final state produced by feedback is not an eigenstate of the measurement operator. This is somewhat analogous to Jacobs' purification speedup protocol, in which the state is maintained to be in an equal superposition of the measurement eigenstates \cite{Jacobs2003}. Application of Jacobs' protocol to an encoded qubit led the authors of \cite{Hill2008} to a method for converting a classically correlated mixed state into a maximally entangled state. The protocol presented here performs the analogous task for an initially pure state. In appendix \ref{sec:FPFDetails}, we show that $\mathcal{C}(t)$ under application of $P_F$ coincides with that given in \cite{Hill2008}. In Sec. \ref{sec:GlobalOptProofs}, we will establish an even more precise connection between $P_F$ and Jacobs' rapid qubit purification protocol. More details are given in appendix \ref{sec:FPFDetails}, where we derive the feedback Hamiltonian that implements $P_F$ and outline $P_F$'s relation to \cite{Hill2008}.
\begin{figure}
\centering
\includegraphics[width = 0.5\textwidth]{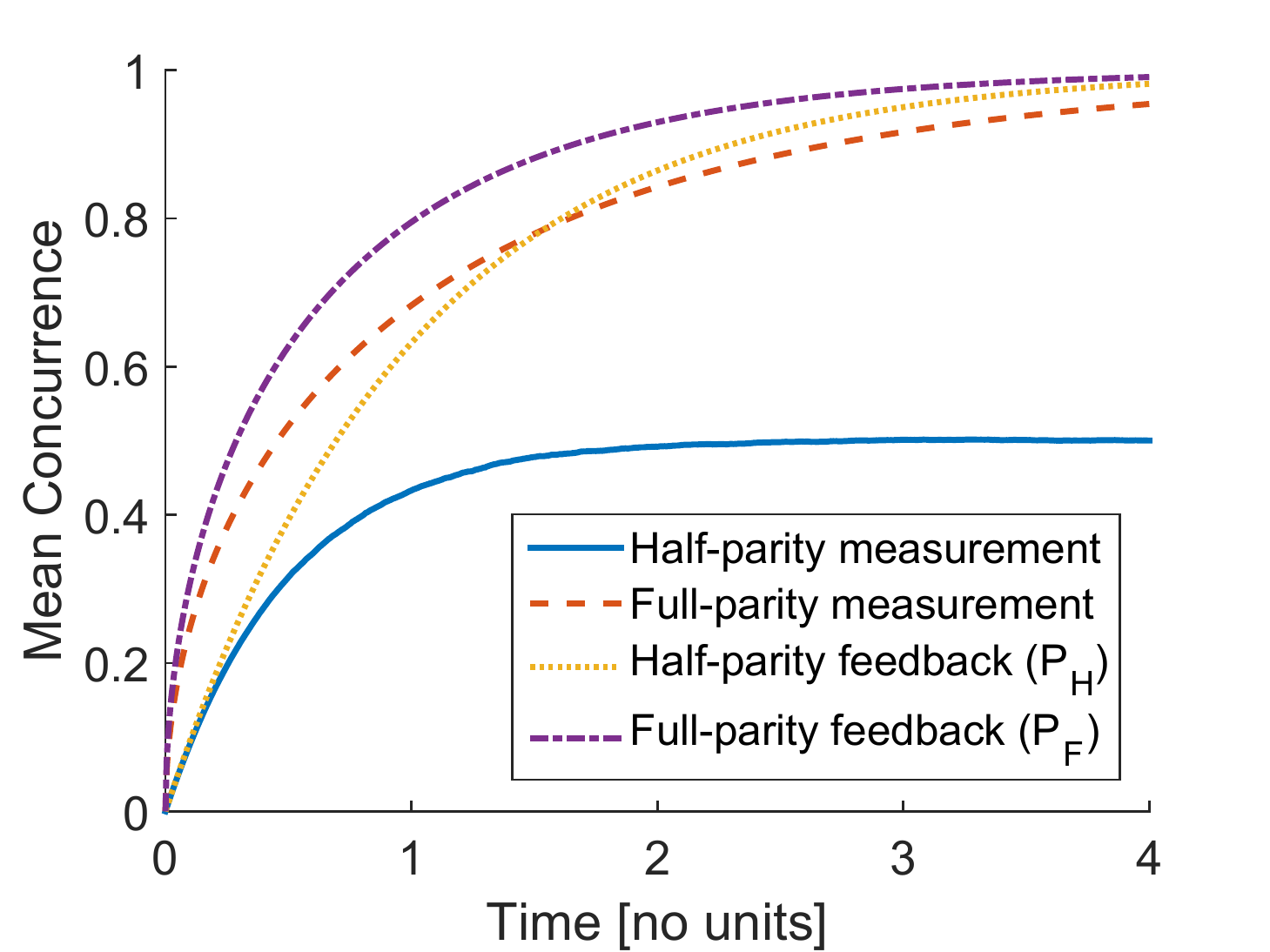}
\caption{(Color online) Expectation value of the concurrence while applying feedback or by applying measurement alone. Using the method of linear quantum trajectories \cite{Jacobs1998linear}, concurrence under full-parity measurement without feedback can be shown to take the form $\mathcal{C}(t) = \text{erf}(\sqrt{t/2})$. Concurrence under half-parity measurement without feedback is calculated numerically by averaging $10000$ trajectories.}
\label{fig:CompareFPF}
\end{figure}

The performance of the half-parity and full-parity feedback protocols are plotted in Fig. \ref{fig:CompareFPF}. We also plot the average concurrence without feedback for both measurement operators. We have normalized the measurement operators $X_F$ and $X_H$ so that the dephasing rate on one of the two qubits (tracing out the other qubit and assuming no feedback) are the same under measurement, allowing quantitative comparison the half- and full-parity performance. It is evident that the full-parity measurement with feedback strictly outperforms all other protocols.
  
\section{Global Optimality In Entanglement Generation}
\label{sec:GlobalOptProofs}

We now set out to prove the optimality results stated in Sec. \ref{sec:Measurement}. For the half-parity feedback protocol, we apply the standard verification theorems often used in control theory to check whether a protocol is globally optimal \cite{Shabani2008, Wiseman2008}. While one might expect that proving such a result would require knowledge about all allowed strategies, it turns out that knowing the performance of the trial protocol is sufficient, so long as one knows how it behaves for any initial state. 

Before using the verification theorems, we first provide a brief summary following Ref. \cite{Shabani2008}. A common feedback goal is to minimize the expectation value of some cost function after applying feedback for a fixed time interval $T$. The cost function can depend on the final state of the system as well as the resources used to apply feedback. From the cost function, one defines the cost-to-go $c(P, \mathbf{x}, t)$ where the vector $\mathbf{x}$ represents the system state at time $t$. Here $c$ is defined to be the expectation value of the cost function at time $T$, assuming that the system was in the state $\mathbf{x}$ at time $t$ and that the controller used feedback protocol $P$ from $t$ to $T$. To check optimality, one first generically writes the equations of motion for $\mathbf{x}$ in terms of their deterministic and stochastic parts as
\begin{align}
\label{eq:dx}
d\mathbf{x} = \mathbf{A}(t, \mathbf{v}_P(t), \mathbf{x}(t))~dt +\mathbf{B}(t, \mathbf{v}_P(t), \mathbf{x}(t))~dW
\end{align}
where $\mathbf{v}_P$ are the feedback control settings and parameters specified by protocol $P$. In the context of quantum feedback, $\mathbf{v}$ can represent Hamiltonian, measurement and dissipation parameters. The procedure can easily be generalized to include multiple noise processes if needed. If $\partial c(P)/\partial t$ and $\partial^2 c(P)/\partial \mathbf{x}^2$ are continuous, then global optimality of $P$ follows from the following two conditions. Firstly, assuming the cost function only depends on the final state, the Hamilton-Jacobi-Bellman equation for the cost-to-go $c(P,\mathbf{x},t)$, given by
\begin{align}
\label{eq:HJB_G}
&G(t, \mathbf{u}, \mathbf{x}, P) \equiv \\ \nn
& -\frac{1}{2} \mathbf{B}(t,\mathbf{u},\mathbf{x})^\intercal \frac{\partial^2 c(P)}{\partial \mathbf{x}^2} \mathbf{B}(t,\mathbf{u},\mathbf{x}) - \mathbf{A}(t,\mathbf{u},\mathbf{x})^\intercal \frac{\partial c(P)}{\partial \mathbf{x}}
\end{align}
and
\begin{align}
\label{eq:HJB}
\frac{\partial c(P)}{\partial t} = \text{max}_{\mathbf{u}} [G(t,\mathbf{u}, \mathbf{x}, P)].
\end{align}
must be satisfied for all time $0$ to $T$ and for all states $\mathbf{x}$. Secondly, $\mathbf{u} = \mathbf{v}_P$ must maximize $G$ for all time and states.

In order to apply the verification theorems, we parameterize the state $|\psi\rangle$ as a vector $\mathbf{x} = \{\mathcal{C}, u, v, w \}$ using \erf{eq:StateParam} and the definitions of $u$, $v$ and $w$ given below \erf{eq:dCHPF2}. Applying a specific local unitary feedback protocol is equivalent to setting $\mathbf{v} = \{u, v, w\}$ equal to specific functions of $t$ and $\mathcal{C}$. 
  
  \textbf{Half-parity measurement: }  For the max. concurrence goal (maximizing $\langle \mathcal{C}(T)\rangle$), we define our cost function to be $1-\mathcal{C}(t)$.
Choosing the locally optimal protocol for the half-parity measurement $P_H: \{u = 0, v = 1, w = 1\}$, we can use the analytic solution \erf{eq:PH} to find the cost-to-go when evolving according to $P_H$ as
\begin{align}
c(P_H, \mathbf{x}, t) = (1-\mathcal{C}(t))e^{t-T}.
\end{align}
The Hamilton-Jacobi-Bellman equation uses the following derivatives of $c$
\begin{align}
\label{eq:HPFPartials}
\frac{\partial c}{\partial t} = (1-\mathcal{C}(t))e^{t-T}, ~~ \frac{\partial c}{\partial\mathbf{x}} = \{-e^{t-T}, 0, 0, 0\}, ~~ \frac{\partial^2 c}{\partial \mathbf{x}^2} = 0
\end{align}
which satisfy the continuity conditions required by the verification theorem. We divide the equations of motion under feedback and half-parity measurement \erf{eq:dCHPF2} into their deterministic and stochastic parts as
\begin{align}
\label{eq:HPFAB}
d\mathcal{C} &= A_\mathcal{C}~dt + B_\mathcal{C}~dW \nn \\
A_\mathcal{C} &= \left\{ \begin{array}{ll}
		[(v^2-u^2)w-\mathcal{C}(v^2+u^2)]dt 	~~ &|~\mathcal{C}>0 \\
		|v^2-u^2|~dt	&|~\mathcal{C}=0
\end{array}
\right. \\ \nn
B_\mathcal{C} &= \left\{ \begin{array}{ll}
		2\mathcal{C}\sqrt{1-\mathcal{C}^2} u v~dW 	~~ &|~\mathcal{C}>0 \\
		0	&|~\mathcal{C}=0
\end{array}
\right. \\ \nn
\end{align}
where we have only calculated the $\mathcal{C}$ components of $\mathbf{A}$ and $\mathbf{B}$. Due to the form of $G$ in \erf{eq:HJB_G}, the other components are unnecessary, since $\partial c/\partial \mathbf{v} = 0$, \textit{i.e.}, our cost function is invariant under the feedback control settings. Substituting Eqs. (\ref{eq:HPFPartials}) and (\ref{eq:HPFAB}) into Eqs. (\ref{eq:HJB_G}) and (\ref{eq:HJB}), we find the condition for $P_H$ to be globally optimal is that it satisfies the maximization condition
\begin{align}
(1-&\mathcal{C}(t))e^{t-T}  \\ \nn
 &= \text{max}_{\{u, v, w\}} [(v^2-u^2)w - \mathcal{C}(t)(v^2+u^2)]e^{t-T}.
\end{align}
The maximum occurs for the values of $u$, $v$ and $w$ specified by $P_H$, and the equation then is satisfied. This proves that the half-parity protocol $P_H$ is globally optimal for maximizing the concurrence at fixed time $T$. Because $c(P_H,\mathbf{x},t)$ is linear in $\mathcal{C}$, one can also show that global optimality follows directly from local optimality in this case \cite{Teo2014}, but this proof method is not applicable in general.

A similar calculation can be performed in which we test for global optimality with respect to the min. time goal \cite{Shabani2008} (minimizing the expected time at which $\mathcal{C}$ reaches some desired value $\mathcal{C}_\text{Threshold}$, called the expected hitting time). One finds that $P_H$ maximizes G only when $\mathcal{C} \leq 1/\sqrt{2}$, which implies that $P_H$ is not globally optimal in general. If one chooses $\mathcal{C}_\text{Threshold}\leq 1/\sqrt{2}$ however, then only the dynamics of the system when $\mathcal{C}\leq 1/\sqrt{2}$ are relevant for determining the expected hitting time. Thus we can instead ask if $P_H$ is globally optimal within the constraint that $\mathcal{C}\leq 1/\sqrt{2}$. Restricted to this parameter space, the Hamilton-Jacobi-Bellman equation Eqs. \ref{eq:HJB_G}-\ref{eq:HJB} is satisfied for all allowed values of $\mathbf{x}$, and $P_H$ maximizes $G$. This proves that $P_H$ is globally optimal for the min. time goal when $\mathcal{C}_\text{Threshold}\leq 1/\sqrt{2}$. As the expected hitting time is the same whether one considers $\mathcal{C}$ or some arbitrary monotonic function $f(\mathcal{C})$ and the corresponding threshold $f(\mathcal{C}_\text{Threshold})$, the min. time proof applies to all monotonic functions of $\mathcal{C}$, such as entanglement of formation.

Concurrence has the favorable property of being invariant under the control parameters, which makes it amenable to methods for proving global optimality. For many tasks however, a specific target state is desired, in which case fidelity is a more relevant figure of merit. We now show that the global optimality proofs given above extend to the corresponding fidelity goals. 

We take any maximally entangled state $|\Psi\rangle$ to be the target state and define $P_H^\Psi$ to be a variant of $P_H$ which rotates $\psi$ to have the maximal fidelity with respect to $\Psi$ at the final time. For the max. fidelity goal, the final time is simply $T$. For the fidelity min. time goal, the final time is the earliest time at which $\mathcal{F}_\text{Threshold}$ can be reached. For both of these goals, we need to know the fidelity of a given state $\psi$ maximized over all local unitaries. For pure states, this maximal fidelity is uniquely determined by the concurrence of $\psi$, and is given by \cite{Verstraete2002fidelity}
\begin{align}
\label{eq:FidConcRelation}
\mathcal{F}_\text{max} = \frac{\mathcal{C}+1}{2}.
\end{align}
In the context of entanglement distillation, $\mathcal{F}_\text{max}$ is often called the singlet fraction, and measures the usefulness of a general quantum state for the task.

To prove that $P_H$ is globally optimal for the fidelity min. time goal, we note that any optimal protocol must have $\mathcal{F} = \mathcal{F}_\text{max}(\mathcal{C})$ at the hitting time. If this were not the case, the protocol could have applied local unitaries to $\psi$ at an earlier time that increase $\mathcal{F}(\psi)$ to $\mathcal{F}_\text{max}(\mathcal{C}(\psi))$, and hence achieve an earlier hitting time. As $\mathcal{F}_\text{max}$ is a monotonic function of $\mathcal{C}$, this relation suffices to prove optimality of $P_H$ with respect to the fidelity min. time goal when $\mathcal{F}_\text{Threshold} \leq \mathcal{F}_\text{max}(1/\sqrt{2}) = (1+\sqrt{2})/\sqrt{8}$. 
  
To prove global optimality of the max. fidelity goal, we assume the evolution of the system is well-approximated by a discrete protocol in which one measures for a small but finite duration before applying feedback \cite{Teo2014}. As realistic implementations of feedback inevitably suffer from feedback delay and finite bandwidth effects at sufficiently short timescales, the continuum limit may not be a good physical model, and we shall therefore not focus on this here. In the discrete approximation, we can express the expectation value of the fidelity at the stopping time $T$ as an integral over all possible measurement outcomes $\{\mathbf{V}\}$. Suppose that some hypothetical feedback protocol $P'$ is globally optimal for the max. fidelity goal. We write
\begin{align}
\label{eq:FidProof1}
\langle \mathcal{F}(P', T)\rangle &= \int \mathcal{F}_\mathbf{V}(P', T) p(\mathbf{V}) d\mathbf{V} \nn \\
 &= \int \mathcal{F}_\text{max}(\mathcal{C}_\mathbf{V}(P', T)) p(\mathbf{V}) d\mathbf{V}
\end{align}
where $\mathcal{F}_\mathbf{V}(P', T)$ and $\mathcal{C}_\mathbf{V}(P', T)$ are respectively the fidelity and concurrence at time $T$, assuming protocol $P'$ was applied and measurement outcome $\mathbf{V}$ occurred. $p(\mathbf{V})$ is the probability of $\mathbf{V}$ occurring (note that $p$ implicitly depends on $P'$ and the state evolution, but this is of no consequence for the proof). The last equality follows from the fact that the globally optimal protocol could perform at least as well if it maximizes the fidelity of each possible final state. If it does not do so on a set of non-zero measure, then a better protocol exists which does. We continue by relating the performance of $P'$ to that of $P_H^\Psi$:
\begin{align}
\label{eq:FidProof2}
 &= \int \frac{\mathcal{C}_\mathbf{V}(P', T)+1}{2} p(\mathbf{V}) d\mathbf{V} \nn \\
 &\leq \int \frac{\mathcal{C}_\mathbf{V}(P_H^\Psi, T)+1}{2} p(\mathbf{V}) d\mathbf{V} \nn \\
 &= \int \mathcal{F}_\text{max}(\mathcal{C}_\mathbf{V}(P_H^\Psi, T)) p(\mathbf{V}) d\mathbf{V} = \int \mathcal{F}_\mathbf{V}(P_H^\Psi, T) p(\mathbf{V}) d\mathbf{V} \nn \\
 &= \langle \mathcal{F}(P_H^\Psi, T)\rangle.
\end{align}
The inequality follows from global optimality of $P_H$ with respect to the max. concurrence goal. Note that relating the performance of $P'$ to that of $P_H$ relies on the fact that $\mathcal{F}_\text{max}$ is linear in $\mathcal{C}$. The second-to-last equality follows from the fact that at time $T$, $P_H^\Psi$ rotates $\psi$ to have maximum fidelity with respect to the target state. Together Eqs. (\ref{eq:FidProof1}) and (\ref{eq:FidProof2}) imply that $P_H^\Psi$ performs at least as well as any potential protocol $P'$, and therefore that $P_H^\Psi$ is globally optimal for the task. 

\textbf{Full-parity measurement: } A connection between Jacobs' rapid purification protocol, and rapid entanglement using full-parity measurement was first established in Ref. \cite{Hill2008}. To prove optimality of the full-parity protocol $P_F$, we observe that the dynamics of the concurrence of a two-qubit pure state under full-parity measurement and local feedback are precisely those of the Bloch vector length of a single continuously monitored qubit with feedback. We then use existing optimality results regarding rapid qubit purification \cite{Wiseman2008} to prove the analogous two-qubit results. 

To see the correspondence between 2-qubit concurrence $\mathcal{C}$ and the Bloch vector length of a single qubit, consider a qubit undergoing a continuous measurement of $\tilde{\sigma}_z$ at a rate $\tilde{\Gamma}$. We assume that some feedback controller can instantly apply any unitary operation at any time, as we have assumed for the two-qubit case. Parameterizing the qubit as a Bloch vector in spherical coordinates $\{\tilde{r}, \tilde{\theta}, \tilde{\phi}\}$, one can use \erf{eq:RhoMCont} to derive an equation of motion for the Bloch vector length $\tilde{r}$ as a function of $\tilde{\theta}$ and $\tilde{\phi}$ \cite{Li2013}
\begin{align}
\label{eq:OneQubit}
d\tilde{r} = (1-\tilde{r}^2) \Big[ \frac{\tilde{\Gamma}}{4\tilde{r}}(1-\tilde{u}^2)dt + \sqrt{\frac{\tilde{\Gamma}}{2}} \tilde{u} ~dW \Big]
\end{align}
where $\tilde{u} = \cos(\tilde{\theta})$ and $\tilde{\theta}$ is the angle the Bloch vector makes with the measurement axis $\tilde{\sigma_z}$. We do not derive equations of motion for $\tilde{\theta}$ and $\tilde{\phi}$ because we assume feedback can set them to their desired values at any time. Making the following identifications
\begin{align}
\label{eq:FPFMapping}
\mathcal{C} &\rightarrow \tilde{r} \nn \\
w &\rightarrow \tilde{u} \nn \\
2(u^2-v^2)^2 &\rightarrow \tilde{\Gamma},
\end{align}
the equation of motion for concurrence under full-parity measurement and feedback, \erf{eq:EOMsF} becomes \erf{eq:OneQubit}. This mapping reveals several interesting features of the dynamics.
As observed in Ref. \cite{Hill2008}, entanglement generation can be turned on and off by rotating the system into one of the decoherence free subspaces of the measurement operator $X_F$. This is evident from the dependence of $\tilde{\Gamma}$ on $u$ and $v$, as $u=v=1$ and $u=v=0$ correspond to states fully localized to the even and odd parity subspaces, respectively. Of particular interest is the direct mapping between $\mathcal{C}$ and $\tilde{r}$; the concurrence coincides exactly with Bloch vector length of the effective qubit. We provide more details on this effective qubit in the appendix.

Although the effective qubit we consider here is different from that discussed in Ref. \cite{Hill2008}, rapid entanglement generation corresponds to applying Jacobs' protocol to the effective qubit in both mappings. $\tilde{u}=0$ is globally optimal for maximizing the linear entropy $\langle \tilde{r}(T) \rangle$ as shown in \cite{Wiseman2008}. Although that work did not consider allowing the measurement rate $\tilde{\Gamma} \leq 2$ to vary, it is straightforward to extend the proofs of \cite{Wiseman2008} to this case by repeating their calculation with $\tilde{\Gamma}$ as a control variable bounded from $0$ to $2$. 
Thus the mapping between equations of motion for $\tilde{r}$ and $\mathcal{C}$ implies that $P_F$ is globally optimal for the max. concurrence goal.

It is also the case that not applying feedback to the effective qubit yields the same equations of motion as not applying feedback to the two-qubits. Since it was proved in \cite{Wiseman2008} that not applying feedback is globally optimal for the min. time goal for linear entropy, the analogous result applies to the full-parity measurement of two qubits for the concurrence min. time goal. The arguments used to extend the max. concurrence and the concurrence min. time results of $P_H$ to the corresponding fidelity results (\textit{i.e.} Eqs. \ref{eq:FidConcRelation}-\ref{eq:FidProof2}) also apply to $P_F$ without modification. This completes the proof of the enumerated results given in Sec. \ref{sec:Measurement}.

\begin{figure}
\centering
\includegraphics[width = 0.5\textwidth]{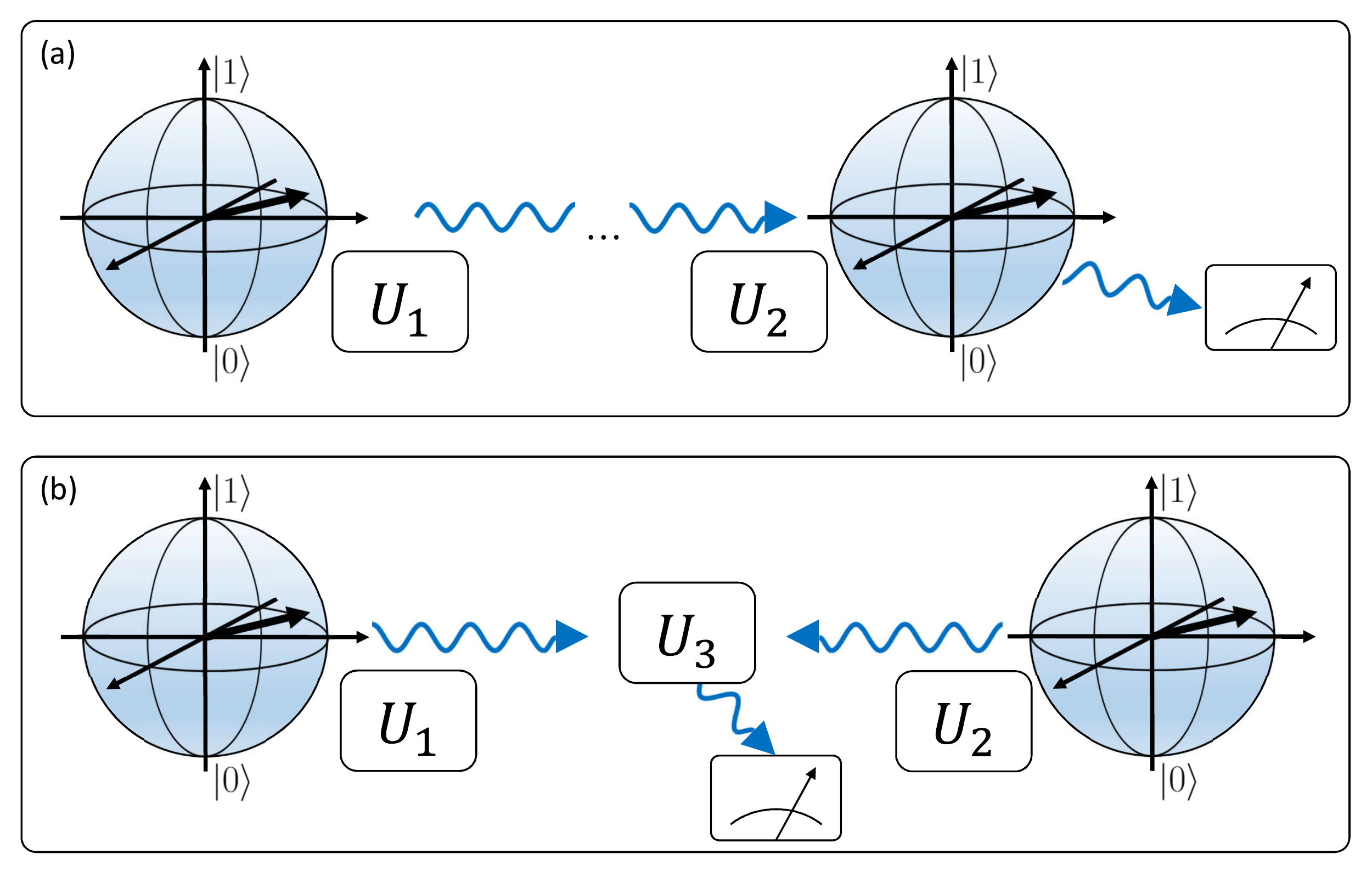}
\caption{(Color online) Remote 2-qubit measurement implementations. (a) Remote 2-qubit measurement in which the signal degrees of freedom (wavey line) propagate from qubit 1 to qubit 2. $U_1$ and $U_2$ represent the interactions between the signal and each qubit, while the dial represents projective measurement of the signal after qubit interactions. These three operations determine the effective 2-qubit measurement operator. (b) Remote measurement based on entanglement swapping. $U_3$ is the interaction between the incoming signals (such as a beam splitter) that erases which-path information. Note that in both schemes, the rate of entanglement generation is bounded by $U_1$.}
\label{fig:RemoteEntanglement}
\end{figure}

\textbf{Upper-bound on measurement-based protocols:} So far, we have focused on optimality given a fixed measurement operator. However, one may ask whether a different measurement operator could offer superior performance using similar resources. In the context of remote entanglement generation, when entanglement is created using some signal degree of freedom as an intermediary (see Fig. \ref{fig:RemoteEntanglement}), one could ask whether entanglement is transferred with unit efficiency.
Motivated by these questions, we now prove a more general result which sets an upper limit on the entanglement entropy achievable generally under a much larger class of measurement-based protocols. 
We will find that the bound is only saturated by the full-parity feedback protocol $P_F$. 

We consider any system in which the action of the measurement on qubit 1 is of the form
\begin{align}
\label{eq:MERho1}
d\rho_1 = (\Gamma_\text{deph.}/2) \mathcal{D}[\vec{\sigma}\cdot \hat{n}]\rho_1~dt
\end{align}
where $\rho_1$ is the state of qubit 1 unconditioned on the measurement outcome and tracing out qubit 2. Physically, the dephasing rate $\Gamma_\text{deph.}$ sets an upper bound on the amount of information that can be extracted from the measurement \cite{Clerk:2010dh}, so this restriction fixes the rate at which information about qubit 1 is transferred to the rest of the system. Thus $\Gamma_\text{deph.}$ defines a physically meaningful reference that lets us compare the performance of $P_H$ and $P_F$ to more general protocols. By tracing out qubit 2 in \erf{eq:RhoMCont}, one arrives at \erf{eq:MERho1} with $\Gamma_\text{deph.} = 1/2$ for measurement of both $X_H$ and $X_F$. 

The entanglement entropy of a state is given by $E_1 = \text{Tr}[\rho_1 \log_2(\rho_1)]$, the Von Neumann entropy of qubit 1. Note that if the joint state is pure, then this expression also equals the entanglement of formation. In order for the entanglement to increase, the entropy of the subsystem must increase, as it does under the action of \erf{eq:MERho1}. As measurement constitutes the only available non-local interaction, \erf{eq:MERho1} fully determines how entanglement may change. At time $t$, the increase in entropy of $\rho_1$ is maximized if the state is unbiased with respect to the measurement axis $\vec{\sigma}\cdot \hat{n}$. For example, if $\vec{\sigma}\cdot \hat{n}=\sigma_z$, then optimal states would be of the form $\rho_1 = (x \sigma_x + y \sigma_y + \sigma_0)/2$ where $\sigma_0$ is the identity matrix. Assuming this condition is satisfied at all times, the entropy as a function of time may be derived by solving \erf{eq:MERho1} with the initial condition $\rho_1 = (|0\rangle + |1\rangle)(\langle 0| + \langle 1|)/2$. Taking $\Gamma_\text{deph.} = 1/2$, we find
\begin{align}
\label{eq:E}
E_1(t) =& -\frac{1-\langle \sigma_{x1} \rangle}{2}\log_2\Big[\frac{1-\langle \sigma_{x1} \rangle}{2} \Big]  \\ &-\frac{1+\langle \sigma_{x1} \rangle}{2}\log_2\Big[\frac{1+\langle \sigma_{x1} \rangle}{2}\Big] \nn \\
\langle \sigma_{x1} \rangle &= e^{-t/2}.
\end{align}
This Von Neumann entropy sets an upper bound on the entanglement entropy that can be achieved by the action of a given measurement operator (as well as on the entanglement of formation if the joint evolution is pure). 

The main result of this section is that by tracing out qubit 2 from 
\erf{eq:StateParam} and substituting \erf{eq:PF},
one can show that the entanglement entropy of qubit 1 is exactly \erf{eq:E}. For pure states, a similar result can be derived for concurrence, which has a one-to-one relation with entanglement entropy in this case \cite{Wootters1998}.
Thus $P_F$ saturates the  bound \erf{eq:E} given by the dephasing rate of $X_F$, and hence no measurement-based protocol of the form \erf{eq:MERho1} can generate entanglement faster than $P_F$. It can also be shown that $P_H$ does not saturate this bound, nor does measurement of $X_F$ or $X_H$ without feedback. 

This bound may be intuitively be understood by considering a remote implementation of the measurement, as depicted in Fig. \ref{fig:RemoteEntanglement}.
\erf{eq:MERho1} then governs how much entanglement is generated between qubit 1 and the ancilla. An ideal entangling protocol would transfer all of this entanglement to qubit 2, so that measurement of the ancilla does not decrease the entanglement entropy. Evidently only $P_F$ fully transfers the entanglement from the measurement signal to qubit 2. For the other protocols, the remaining entanglement is destroyed when the signal is measured. 

Note that if the effect of measurement on qubit 1 yields a dephasing operator that is not normal (\textit{i.e.}, $[X,X^\dagger]\neq0$), then it cannot be put in the form of \erf{eq:MERho1} and our derivation does not apply. 
Thus our bound does not apply for quantum-demolition measurements such as a spontaneous emission process on the first qubit \textit{i.e.}, $d\rho_1 = \mathcal{D}[\sigma]\rho_1~dt$, though a similar bound could be derived for such cases. An example of the latter is entanglement generation via Hong-Ou-Mandel interference.

\section{Conclusion}
\label{sec:Conclusion}

We have derived equations of motion for the concurrence of a two-qubit state under non-local measurement and arbitrary local feedback control. These equations provide a simple and general method for deriving feedback protocols, allowing us to rederive a feedback protocol for deterministic entanglement generation using a half-parity measurement, and to derive a protocol for faster entanglement of initially separable pure states using a full-parity measurement. Parameterization of the state in terms of an entanglement monotone and local operations has allowed us to prove several global optimality results in both cases. In studying the full-parity protocol, our work elaborates on the connections between rapid purification and rapid entanglement generation of Ref. \cite{Hill2008} by quantitatively linking the dynamics of the Bloch vector length to that of the concurrence. We have also derived a bound on rapid entanglement generation given by the dephasing of a single measurement, which is saturated only by the full-parity feedback protocol, indicating that it transfers entanglement with unit efficiency.

Our work suggests several general avenues for further study. Regarding the mapping between a two-qubit pure state and a one qubit mixed state, it is unclear if this connection is an isolated case or if it is indicative of a more general set of mappings between pure states and mixed states of smaller systems. If such mappings did exist, they could potentially simplify the analysis of multi-partite entangled systems. Several further open questions relate to entanglement generation and feedback. Our optimality results for the min. time goals involving the half-parity measurement only apply when the target concurrence is less than $1/\sqrt{2}$. Optimal protocols for the remaining parameter regime could be found by numerical searches. 
Our work could also be extended to consider optimality under many other goals and cost functions, such as entanglement of formation or concurrence of the ensemble density matrix unconditioned on the measurement record.

There are also numerous possible extensions and variations to the physical systems that we have considered so far. The full-parity scheme generates entanglement very rapidly at short times, but only reaches unit concurrence asymptotically, unlike a simple CNOT gate. Combining entangling unitaries of finite strength with measurement and feedback might yield even faster entanglement generation. Parity measurements can also be useful for generating a variety of entangled states, from the three-particle GHZ and W states to entangled many-body states such as cluster states and the Surface code ground state \cite{Fowler2012surface}.
Feedback could be used to generate such states more efficiently and perhaps optimally. 

Furthermore, considerable experimental and theoretical work has focused on the task of generating non-classical states using dissipation alone \cite{Shankar2013autonomously, Schwartz2015, Motzoi2015back, ticozzi2009analysis, ticozzi2012stabilizing, JohnsonViola2015}. Given a fixed coupling strength to the environment, one could ask what dissipation channels are optimal for entanglement generation, or how to apply feedback if the dissipation channels are monitored \cite{liu2016comparing}. The latter is particularly amenable to our techniques, since it could involve pure-state dynamics.

Because the measurements we have studied are candidates for remote entanglement generation, studying optimality in the presence of loss remains a crucial detail to be considered in future work. In this context, entanglement distillation will often be required \cite{Wootters1997}, and will determine which cost functions are relevant in practice. For example, given a specific distillation protocol, one could ask to generate a distilled qubit with sufficiently high concurrence in as little time as possible, which would in turn dictate the cost function to use when optimizing the entanglement generation process. Other natural considerations include other experimental defects such as finite qubit coherence, feedback delay and finite measurement efficiency.

\begin{acknowledgments}
We thank Hanhan Li and Felix Motzoi for their comments on an early draft of this manuscript. LM was supported by the National Science Foundation Graduate Fellowship Grant No. (1106400) and the Berkeley Fellowship for Graduate Study. This work was supported by Laboratory Directed Research and Development (LDRD) funding from Lawrence Berkeley National Laboratory, provided by the U.S. Department of Energy, Office of Science under Contract No. DE-AC02-05CH11231.
\end{acknowledgments}

\appendix
\section{Details on the derivations of Eqs. \ref{eq:dCHPF2} and \ref{eq:EOMsF}}
\label{sec:CalcDetails}

To derive \erf{eq:dCHPF2}, we use \erf{eq:Concurrence} and divide the quantity inside the absolute value into three terms, the initial concurrence $\mathcal{C}(t)$, the change in the real part $d\mathcal{C}_\mathbb{R}$ and the change in the imaginary part $d\mathcal{C}_\mathbb{I}$. We have chosen the global phase of $\psi$ in \erf{eq:StateParam} so that $\mathcal{C}(t)$ appears in the expression without a phase factor.

\begin{align}
\label{eq:dCComplex}
\mathcal{C}&(t+dt) = \Big{|}\langle \psi(t+dt)^* | \sigma_y \otimes \sigma_y |\psi(t+dt)\rangle \Big{|} \\ \nn
&= |\mathcal{C}(t) + d\mathcal{C}_\mathbb{R} + i d\mathcal{C}_\mathbb{I}| \\ \nn
d\mathcal{C}_\mathbb{R} &= 2\mathcal{C}(t)\sqrt{1-\mathcal{C}(t)^2}u v ~dW \\ \nn &+ [(v^2 - u^2)\cos(2\phi) - \mathcal{C}(t)(v^2 + u^2)]~dt \\ \nn
d\mathcal{C}_\mathbb{I} &= \sqrt{1-\mathcal{C}(t)^2}(v^2 - u^2)\sin(2 \phi)~dt
\end{align}
If $\mathcal{C}\neq0$, we can use Ito's lemma and the standard relation $|z|=\sqrt{\text{Re}[z]^2+\text{Im}[z]^2}$ to simplify the above expression.
\begin{align}
d\mathcal{C} = &2\mathcal{C}\sqrt{1-\mathcal{C}^2} u v~dW \\ \nn
 &+ [(v^2-u^2)w-\mathcal{C}(v^2+u^2)]dt 	~~~~~~\mathcal{C}>0
\end{align}

The $\mathcal{C}=0$ case must be treated separately, as $|z|$ is not twice-differentiable at the origin as required by Ito's lemma. At $\mathcal{C}=0$, $\psi$ is independent of the value of $\phi$ up to a global phase. Thus the evolution of $\mathcal{C}$ here cannot be affected by the value of $\phi$, and hence we can restrict the value of the latter without loss of generality.
We can avoid taking the absolute value and directly assign $d\mathcal{C}|_{\mathcal{C}=0} = d\mathcal{C}_\mathbb{R}$ by enforcing $\sin(2\phi)=0$, so that $d\mathcal{C}_\mathbb{I}=0$. To keep $d\mathcal{C}$ non-negative, we choose $\phi$ such that $\cos(2 \phi)=1$ if $(v^2-u^2)>0$ and $\cos(2\phi)=-1$ otherwise, which yields
\begin{equation}
d\mathcal{C} = |v^2-u^2|~dt ~~~~~~\mathcal{C}=0.
\end{equation}
This completes the derivation of \erf{eq:dCHPF2}.

 There is one subtle difference between the derivations of \erf{eq:EOMsF} and \erf{eq:dCHPF2},
 which accounts for the fact that we have allowed $\mathcal{C}$ to take on negative values in \erf{eq:EOMsF}.
In the full-parity case, $d\mathcal{C}_{\mathbb{R}/\mathbb{I}}|_{\mathcal{C}=0}$ depend on $dW$, so it is not possible to choose $\phi$ to maintain $d\mathcal{C}_\mathbb{R}\geq0$. Even mathematically it does not make sense to choose $\phi$ to anticipate the value of $dW$, as the resulting integrand would not be an adapted process, and hence the Ito integral would not be defined \cite{Oksendal2003}. 
To remedy this technicality, we set $\phi|_{\mathcal{C}=0}=0$ and allow $\mathcal{C}$ to take on negative values (as before, $\phi=0$ sets $d\mathcal{C}_\mathbb{I}=0$). As $\psi(\mathcal{C})=\psi(-\mathcal{C})$, one may interpret a state with $\mathcal{C}<0$ as having concurrence $|\mathcal{C}|$. This step yields the $\mathcal{C}=0$ case of \erf{eq:EOMsF}. To make sure the equations of motion for $\mathcal{C}$ are consistent with the underlying equations of motion for $\psi$, we take $\mathcal{C}(t+dt) = \text{sgn}(\mathcal{C}(t))|\mathcal{C}(t) + d\mathcal{C}_\mathbb{R} + id \mathcal{C}_\mathbb{I}|$ when calculating the $\mathcal{C}\neq 0$ case of \erf{eq:EOMsF}, where $\text{sgn}(x)$ is the sign function, defined to be $\pm1$ for $x \gtrless 0$ and 0 otherwise. 

\section{Feedback Hamiltonian for the full-parity protocol and its relation to that of Hill \textit{et al.}\cite{Hill2008}}
\label{sec:FPFDetails}

In section \ref{sec:Feedback}, we derived a protocol for rapid entanglement generation using local unitary feedback and a full-parity measurement. This protocol is quite similar to that given in Ref. \cite{Hill2008}, but turns out to be distinct from it in several important respects. In this section, we first derive how to implement our protocol using Hamiltonian feedback, and then outline the similarities and differences between it and the protocol of Ref. \cite{Hill2008}.

Using the Schmidt decomposition \erf{eq:StateParam}, $P_F$ specifies the state evolution $\psi(t)$ without giving the Hamiltonian feedback necessary to induce this particular evolution. We now use $\psi(t)$ to derive the required Hamiltonian. We assume as is often the case that the protocol can be implemented using proportional feedback, meaning that the controller applies the Hamiltonian $H_F ~dV$. In this case, the equation of motion for the system is the Wiseman-Milburn feedback master equation \cite{Wiseman2009}
\begin{align}
\label{eq:WisemanMilburn}
d\psi  =& \Big[\frac{-1}{2} \big(X_F-\langle X_F \rangle\big)^2 dt + \big(X_F-\langle X_F\rangle\big) dW  \\
& + \Big[\frac{-i}{2}H_F \big(X_F+\langle X_F\rangle\big) - \frac{H_F^2}{8}\Big]dt - \frac{i}{2} H_F ~dW \Big] \psi \nn
\end{align}
%
where we have set $\hbar = 1$. The first two terms match \erf{eq:PsiMCont} and correspond to continuous measurement of $X_F$. The remaining terms give the evolution under feedback. We know from the derivation of $P_F$ that $H_F$ consists entirely of local rotations if it exists (see Sec. \ref{sec:Feedback}). To calculate $H_F$, we first express it in the Pauli basis
\begin{align}
\label{eq:HFPauli}
H_F =  h_0 I + \sum_{i=\{1,2\}} h_{x, i} \sigma_{x,i} + h_{y, i} \sigma_{y,i} + h_{z, i} \sigma_{z,i}
\end{align}
where $I$ is the identity operator. We have added the physically meaningless constant $h_0$ fit the evolution of the global phase of $\psi(t)$, which is already implicitly specified. Substituting \erf{eq:HFPauli} and $\psi(t)$ given by \erf{eq:FPFPsiSol} into \erf{eq:WisemanMilburn}, we solve for $h_{i, j}$. The calculation is simplified by the fact that $\langle X_F \rangle = 0$, and may be further simplified by taking advantage of the fact that $\psi(t)$ evolves deterministically, and hence the $dW$ terms of \erf{eq:WisemanMilburn} must cancel. Physically, this means that the feedback Hamiltonian is chosen to exactly cancel the stochastic back-action from the measurement. Thus we begin by finding $h_{i,j}$ such that
\begin{align}
\label{eq:DetSol}
\Big(X_F - \frac{i}{2}H_F \Big)\psi(t) = 0.
\end{align}
For $\Delta \gamma = \gamma = 0$, $u = 0$ and $v = 1$, \erf{eq:DetSol} admits the solution
\begin{align}
\label{eq:HFSol}
h_{x,1} + h_{x, 2} &= \frac{-1}{|\mathcal{C}|}, ~ h_{y, i} = h_{z, i} = 0 \nn \\
h_0 &= -\frac{1-\mathcal{C}^2}{|\mathcal{C}|}.
\end{align}
Solutions for other values of $\gamma$ and $\Delta \gamma$ may be obtained by a local change of basis of $H_F$. Substituting the solution for $\mathcal{C}(t)$ of \erf{eq:PF} into \erf{eq:HFSol}, it can be verified that the $dt$ terms of \erf{eq:WisemanMilburn} are also satisfied by \erf{eq:HFSol}. Thus for $\mathcal{C} \neq 0$, $P_F$ may be implemented by applying local $\sigma_x$ rotations by an angle proportional to the measurement outcome $dV$. Notice that from \erf{eq:StateParam}, this is about an axis orthogonal to that of the $\pi/2$ pulses required to prepare the initial state assuming one starts in the ground state $|11\rangle$. Also note that $h_{x,1} = 0$, $h_{x,2} = 0$ and $h_{x,1} = h_{x,2} = 1/2\mathcal{C}$ are all solutions, indicating that one may implement $P_F$ by applying rotations on only one of the two qubits, or by applying the same rotation to both. The former is useful for remote entanglement; after the measurement signal interacts with qubit 2 and is read out by the controller, the controller can apply feedback only to qubit 2, eliminating the detrimental effect of propagation delay from qubit 1 to the controller.

For $\mathcal{C}=0$, feedback apparently requires pulses that rotate the qubits by a finite angle. The fact that $\langle X_F \rangle = 0$ allows us to determine these pulses. It can be shown that any $\sigma_x$ rotations satisfying $\theta_1 + \theta_2 = -\text{sgn}(dV) \pi/2$ induce the desired evolution. Comparing with Jacobs' rapid qubit purification protocol \cite{Jacobs2003}, we see that the feedback for $P_F$ takes exactly the same form.

$P_F$ bears many similarities to the protocol of Ref. \cite{Hill2008}. Both involve applying Jacobs' protocol to an encoded qubit, and the concurrence as a function of time is the same. Furthermore, they can both be implemented using only single qubit rotations. $P_F$ is defined using the pure state Schmidt decomposition, and therefore is not defined for mixed states. The protocol of Ref. \cite{Hill2008} is defined by applying Jacobs' purification protocol to an effective qubit encoded via the mapping
\begin{align}
\label{eq:HillQubit}
&\sigma_x' \rightarrow  \sigma_{x2} \nn \\ 
&\sigma_y' \rightarrow  \sigma_{z1}\sigma_{y2} \nn \\ 
&\sigma_z' \rightarrow  \sigma_{z1}\sigma_{z2}.
\end{align}
The authors apply this protocol to a classically correlated mixture, which simultaneously purifies and entangles the qubits. To compare with $P_F$, we substitute \erf{eq:PF} into \erf{eq:HillQubit}
\begin{align}
&\langle \sigma_x' \rangle = -\sqrt{1-\mathcal{C}^2} \cos(\gamma - \Delta \gamma) \nn \\
&\langle \sigma_y' \rangle =  \mathcal{C} \cos(\gamma - \Delta \gamma) \nn \\
&\langle \sigma_z' \rangle = 0.
\end{align}
$P_F$ does indeed maintain $\langle \sigma_z' \rangle = 0$ like Jacobs' rapid purification protocol, but it's action on this effective qubit is not to purify it, but rather to rotate it about the $\sigma_z'$ axis. Its purity remains constant, determined by the initial preparation of the qubits. When $\tilde{\Gamma}=2$, it is natural to consider $\tilde{r}$ of \erf{eq:FPFMapping} to represent two effective qubits encoded as
\begin{align}
\label{eq:Qeff1}
&\langle \tilde{\sigma}_x^1\rangle \rightarrow  -\langle\sigma_{x1}\sigma_{z2}\rangle = \mathcal{C} \sin(\gamma_1)\sin(2\phi) \nn \\ 
&\langle \tilde{\sigma}_y^1\rangle \rightarrow  \langle\sigma_{y1}\sigma_{z2}\rangle = \mathcal{C} \cos(\gamma_1)\sin(2\phi) \nn \\ &\langle \tilde{\sigma}_z^1\rangle \rightarrow  -\langle\sigma_{z1}\sigma_{z2}\rangle = \mathcal{C} \cos(2\phi)
\end{align}
\begin{align}
\label{eq:Qeff2}
&\langle \tilde{\sigma}_x^2\rangle \rightarrow  -\langle\sigma_{z1}\sigma_{x2}\rangle = \mathcal{C} \sin(\gamma_2)\sin(2\phi) \nn \\
&\langle \tilde{\sigma}_y^2\rangle \rightarrow  \langle\sigma_{z1}\sigma_{y2}\rangle = \mathcal{C} \cos(\gamma_2)\sin(2\phi) \nn \\
&\langle \tilde{\sigma}_z^2\rangle \rightarrow  -\langle\sigma_{z1}\sigma_{z2}\rangle = \mathcal{C} \cos(2\phi).
\end{align}
where we have set $\theta=\pi/2$ and $\Delta \theta=0$, although other allowed values yield similar results. Note that the operators on the left-hand side of each equality together obey the Pauli commutation relations $[\sigma_a, \sigma_b] = 2i \epsilon_{abc}\sigma_c$ and therefore represent valid encoded qubits (although they are not independent, as they share the same $z$ component \textit{i.e.}, $\langle\tilde\sigma_z^1\rangle = \langle\tilde\sigma_z^2\rangle$). The right-hand sides are functions of local unitary operations on the physical qubits and coincide with a Bloch sphere representation in spherical coordinates when we make the mapping $\{\mathcal{C}, 2\phi, \gamma_1\}$ or $\{\mathcal{C}, 2\phi, \gamma_2\} \rightarrow \{\tilde{r}, \tilde{\theta}, \tilde{\phi}\}$.

To conclude, we show that even though the protocols involve different initial states, the performance as measured by the concurrence is equivalent. This equivalence is somewhat non-trivial since the expression for the concurrence of mixed states is a non-trivial generalization of that for pure states. Our analysis also uncovers further explanation for the connection between rapid purification and rapid entanglement generation.

We begin again with the Wiseman-Milburn equation for unit efficiency measurement, this time written as a master equation
\begin{align}
\label{eq:WMSME}
d\rho =& \mathcal{D}[X_F]\rho dt + \mathcal{H}[X_F]\rho dW - i[H_F,\rho]\frac{dW}{2} \\
-&i[H_f,X_F \rho+\rho X_F]\frac{dt}{2}+\mathcal{D}[H_F]\rho \frac{dt}{4}.\nn
\end{align}
We follow the prescription given in \cite{Hill2008}, starting with the mixed state $\rho_0 = (|00\rangle \langle 00| + |11\rangle \langle 11|)/2$. This state lies within a decoherence-free subspace of $X_F$, so Hadamard rotations are first applied to both qubits so that the measurement acts non-trivially on the state. Jacobs' rapid purification protocol is then applied on an encoded qubit by applying local $\sigma_x$ rotations on qubit 1. For this, we set $H_F = P \sigma_{x,1}/2$ where $P$ is to be determined. Like Jacobs' protocol, the optimal feedback rotation is then determined by requiring $\langle X_F \rangle=0$ after each application of feedback, which keeps the state maximally unbiased with respect to the measurement operator. The resulting evolution preserves spin flip symmetry ($\rho \rightarrow \sigma_y \otimes \sigma_y \rho^* \sigma_y\otimes \sigma_y$), exchange symmetry, and is deterministic, which makes the dynamics easy to find. The state evolution starting in $\rho_0$ may be fully parameterized by one variable as
%
\begin{align}
\label{eq:rhoHillRalph}
\rho(t) = \begin{bmatrix}
1/4 & \alpha(t) & \alpha(t) & 1/4 \\
-\alpha(t) & 1/4 & 1/4 & -\alpha(t) \\
-\alpha(t) & 1/4 & 1/4 & -\alpha(t) \\
1/4 & \alpha(t) & \alpha(t) & 1/4 \end{bmatrix}
\end{align}
with $\alpha(t)$ purely imaginary. Note that from \erf{eq:HillQubit}, one can show that $\alpha(t)$ is proportional to the Bloch vector length of the effective qubit considered in Ref. \cite{Hill2008}. This evolution follows from $P = i/(2\alpha)$, which may be determined by requiring $\langle X_F \rangle = 0$ or by searching for a deterministic solution. Using \erf{eq:WMSME}, one finds the solution $\alpha(t) = -i \sqrt{1-e^{-t}}/4$

Because $\rho(t)$ is spin flip symmetric, its concurrence is directly related to its eigenvalues. Thus purification and entanglement generation are nearly equivalent tasks. If the eigenvalues of $\rho$ are $\{\lambda_1, \lambda_2, \lambda_3, \lambda_4\}$ written in decreasing order, then the concurrence is $\text{max}\{0, \lambda_1-\lambda_2-\lambda_3-\lambda_4\}$ and the purity is $\sum_i \lambda_i^2$. Both functions are large when the magnitude of one eigenvalue dominates, and thus protocols for entanglement and purification should attempt to maximize the one eigenvalue of $\rho$ while suppressing the others. Using the former equation, we find $\mathcal{C}(t) = \sqrt{1-e^{-t}}$, which coincides with the solution \erf{eq:PF} with $\mathcal{C}(0) = 0$.

\bibliography{GlObib02}

\begin{thebibliography}{46}
\expandafter\ifx\csname natexlab\endcsname\relax\def\natexlab#1{#1}\fi
\expandafter\ifx\csname bibnamefont\endcsname\relax
  \def\bibnamefont#1{#1}\fi
\expandafter\ifx\csname bibfnamefont\endcsname\relax
  \def\bibfnamefont#1{#1}\fi
\expandafter\ifx\csname citenamefont\endcsname\relax
  \def\citenamefont#1{#1}\fi
\expandafter\ifx\csname url\endcsname\relax
  \def\url#1{\texttt{#1}}\fi
\expandafter\ifx\csname urlprefix\endcsname\relax\def\urlprefix{URL }\fi
\providecommand{\bibinfo}[2]{#2}
\providecommand{\eprint}[2][]{\url{#2}}

\bibitem[{\citenamefont{Broadbent et~al.}(2009)\citenamefont{Broadbent,
  Fitzsimons, and Kashefi}}]{broadbent2009universal}
\bibinfo{author}{\bibfnamefont{A.}~\bibnamefont{Broadbent}},
  \bibinfo{author}{\bibfnamefont{J.}~\bibnamefont{Fitzsimons}},
  \bibnamefont{and} \bibinfo{author}{\bibfnamefont{E.}~\bibnamefont{Kashefi}},
  in \emph{\bibinfo{booktitle}{Foundations of Computer Science, 2009. FOCS'09.
  50th Annual IEEE Symposium on}} (\bibinfo{organization}{IEEE},
  \bibinfo{year}{2009}), pp. \bibinfo{pages}{517--526}.

\bibitem[{\citenamefont{Fowler et~al.}(2012)\citenamefont{Fowler, Mariantoni,
  Martinis, and Cleland}}]{Fowler2012surface}
\bibinfo{author}{\bibfnamefont{A.~G.} \bibnamefont{Fowler}},
  \bibinfo{author}{\bibfnamefont{M.}~\bibnamefont{Mariantoni}},
  \bibinfo{author}{\bibfnamefont{J.~M.} \bibnamefont{Martinis}},
  \bibnamefont{and} \bibinfo{author}{\bibfnamefont{A.~N.}
  \bibnamefont{Cleland}}, \bibinfo{journal}{Phys. Rev. A}
  \textbf{\bibinfo{volume}{86}}, \bibinfo{pages}{032324}
  (\bibinfo{year}{2012}).

\bibitem[{\citenamefont{Devoret and
  Schoelkopf}(2013)}]{Devoret2013superconducting}
\bibinfo{author}{\bibfnamefont{M.~H.} \bibnamefont{Devoret}} \bibnamefont{and}
  \bibinfo{author}{\bibfnamefont{R.~J.} \bibnamefont{Schoelkopf}},
  \bibinfo{journal}{Science} \textbf{\bibinfo{volume}{339}},
  \bibinfo{pages}{1169} (\bibinfo{year}{2013}).

\bibitem[{\citenamefont{Nickerson et~al.}(2013)\citenamefont{Nickerson, Li, and
  Benjamin}}]{Benjamin2013}
\bibinfo{author}{\bibfnamefont{N.}~\bibnamefont{Nickerson}},
  \bibinfo{author}{\bibfnamefont{Y.}~\bibnamefont{Li}}, \bibnamefont{and}
  \bibinfo{author}{\bibfnamefont{S.}~\bibnamefont{Benjamin}},
  \bibinfo{journal}{Nat. Commun.} \textbf{\bibinfo{volume}{4}},
  \bibinfo{pages}{1756} (\bibinfo{year}{2013}).

\bibitem[{\citenamefont{Motzoi et~al.}(2015)\citenamefont{Motzoi, Whaley, and
  Sarovar}}]{Motzoi2015}
\bibinfo{author}{\bibfnamefont{F.}~\bibnamefont{Motzoi}},
  \bibinfo{author}{\bibfnamefont{K.~B.} \bibnamefont{Whaley}},
  \bibnamefont{and} \bibinfo{author}{\bibfnamefont{M.}~\bibnamefont{Sarovar}},
  \bibinfo{journal}{Phys. Rev. A} \textbf{\bibinfo{volume}{92}},
  \bibinfo{pages}{032308} (\bibinfo{year}{2015}).

\bibitem[{\citenamefont{Roch et~al.}(2014)\citenamefont{Roch, Schwartz, Motzoi,
  Macklin, Vijay, Eddins, Korotkov, Whaley, Sarovar, and Siddiqi}}]{Roch2014}
\bibinfo{author}{\bibfnamefont{N.}~\bibnamefont{Roch}},
  \bibinfo{author}{\bibfnamefont{M.~E.} \bibnamefont{Schwartz}},
  \bibinfo{author}{\bibfnamefont{F.}~\bibnamefont{Motzoi}},
  \bibinfo{author}{\bibfnamefont{C.}~\bibnamefont{Macklin}},
  \bibinfo{author}{\bibfnamefont{R.}~\bibnamefont{Vijay}},
  \bibinfo{author}{\bibfnamefont{A.~W.} \bibnamefont{Eddins}},
  \bibinfo{author}{\bibfnamefont{A.~N.} \bibnamefont{Korotkov}},
  \bibinfo{author}{\bibfnamefont{K.~B.} \bibnamefont{Whaley}},
  \bibinfo{author}{\bibfnamefont{M.}~\bibnamefont{Sarovar}}, \bibnamefont{and}
  \bibinfo{author}{\bibfnamefont{I.}~\bibnamefont{Siddiqi}},
  \bibinfo{journal}{Phys. Rev. Lett.} \textbf{\bibinfo{volume}{112}},
  \bibinfo{pages}{170501} (\bibinfo{year}{2014}).

\bibitem[{\citenamefont{Hofmann et~al.}(2012)\citenamefont{Hofmann, Krug,
  Ortegel, G\'{e}rard, Weber, Rosenfeld, and Weinfurter}}]{Hofmann2012}
\bibinfo{author}{\bibfnamefont{J.}~\bibnamefont{Hofmann}},
  \bibinfo{author}{\bibfnamefont{M.}~\bibnamefont{Krug}},
  \bibinfo{author}{\bibfnamefont{N.}~\bibnamefont{Ortegel}},
  \bibinfo{author}{\bibfnamefont{L.}~\bibnamefont{G\'{e}rard}},
  \bibinfo{author}{\bibfnamefont{M.}~\bibnamefont{Weber}},
  \bibinfo{author}{\bibfnamefont{W.}~\bibnamefont{Rosenfeld}},
  \bibnamefont{and}
  \bibinfo{author}{\bibfnamefont{H.}~\bibnamefont{Weinfurter}},
  \bibinfo{journal}{Science} \textbf{\bibinfo{volume}{337}},
  \bibinfo{pages}{72} (\bibinfo{year}{2012}).

\bibitem[{\citenamefont{Bernien et~al.}(2013)\citenamefont{Bernien, Hensen,
  Pfaff, Koolstra, Blok, Robledo, Taminiau, Markham, Twitchen, Childress
  et~al.}}]{Bernien2013}
\bibinfo{author}{\bibfnamefont{H.}~\bibnamefont{Bernien}},
  \bibinfo{author}{\bibfnamefont{B.}~\bibnamefont{Hensen}},
  \bibinfo{author}{\bibfnamefont{W.}~\bibnamefont{Pfaff}},
  \bibinfo{author}{\bibfnamefont{G.}~\bibnamefont{Koolstra}},
  \bibinfo{author}{\bibfnamefont{M.}~\bibnamefont{Blok}},
  \bibinfo{author}{\bibfnamefont{L.}~\bibnamefont{Robledo}},
  \bibinfo{author}{\bibfnamefont{T.~H.} \bibnamefont{Taminiau}},
  \bibinfo{author}{\bibfnamefont{M.}~\bibnamefont{Markham}},
  \bibinfo{author}{\bibfnamefont{D.}~\bibnamefont{Twitchen}},
  \bibinfo{author}{\bibfnamefont{L.}~\bibnamefont{Childress}},
  \bibnamefont{et~al.}, \bibinfo{journal}{Nature}
  \textbf{\bibinfo{volume}{497}}, \bibinfo{pages}{86} (\bibinfo{year}{2013}).

\bibitem[{\citenamefont{{\v{C}}ernot{\'\i}k and
  Hammerer}(2015)}]{Vcernotik2015measurement}
\bibinfo{author}{\bibfnamefont{O.}~\bibnamefont{{\v{C}}ernot{\'\i}k}}
  \bibnamefont{and} \bibinfo{author}{\bibfnamefont{K.}~\bibnamefont{Hammerer}},
  \bibinfo{journal}{arXiv preprint arXiv:1512.00768}  (\bibinfo{year}{2015}).

\bibitem[{\citenamefont{Roy et~al.}(2015)\citenamefont{Roy, Jiang, Stone, and
  Devoret}}]{Roy2015remote}
\bibinfo{author}{\bibfnamefont{A.}~\bibnamefont{Roy}},
  \bibinfo{author}{\bibfnamefont{L.}~\bibnamefont{Jiang}},
  \bibinfo{author}{\bibfnamefont{A.~D.} \bibnamefont{Stone}}, \bibnamefont{and}
  \bibinfo{author}{\bibfnamefont{M.}~\bibnamefont{Devoret}},
  \bibinfo{journal}{Phys. Rev. Lett.} \textbf{\bibinfo{volume}{115}},
  \bibinfo{pages}{150503} (\bibinfo{year}{2015}).

\bibitem[{\citenamefont{Silveri et~al.}(2016)\citenamefont{Silveri,
  Zalys-Geller, Hatridge, Leghtas, Devoret, and Girvin}}]{Silveri2016theory}
\bibinfo{author}{\bibfnamefont{M.}~\bibnamefont{Silveri}},
  \bibinfo{author}{\bibfnamefont{E.}~\bibnamefont{Zalys-Geller}},
  \bibinfo{author}{\bibfnamefont{M.}~\bibnamefont{Hatridge}},
  \bibinfo{author}{\bibfnamefont{Z.}~\bibnamefont{Leghtas}},
  \bibinfo{author}{\bibfnamefont{M.~H.} \bibnamefont{Devoret}},
  \bibnamefont{and} \bibinfo{author}{\bibfnamefont{S.}~\bibnamefont{Girvin}},
  \bibinfo{journal}{Phys. Rev. A} \textbf{\bibinfo{volume}{93}},
  \bibinfo{pages}{062310} (\bibinfo{year}{2016}).

\bibitem[{\citenamefont{Motzoi et~al.}(2016)\citenamefont{Motzoi, Halperin,
  Wang, Whaley, and Schirmer}}]{Motzoi2015back}
\bibinfo{author}{\bibfnamefont{F.}~\bibnamefont{Motzoi}},
  \bibinfo{author}{\bibfnamefont{E.}~\bibnamefont{Halperin}},
  \bibinfo{author}{\bibfnamefont{X.}~\bibnamefont{Wang}},
  \bibinfo{author}{\bibfnamefont{K.~B.} \bibnamefont{Whaley}},
  \bibnamefont{and} \bibinfo{author}{\bibfnamefont{S.}~\bibnamefont{Schirmer}},
  \bibinfo{journal}{Phys. Rev. A} \textbf{\bibinfo{volume}{94}},
  \bibinfo{pages}{032313} (\bibinfo{year}{2016}).

\bibitem[{\citenamefont{Sarovar et~al.}(2005)\citenamefont{Sarovar, Goan,
  Spiller, and Milburn}}]{Sarovar2005}
\bibinfo{author}{\bibfnamefont{M.}~\bibnamefont{Sarovar}},
  \bibinfo{author}{\bibfnamefont{H.-S.} \bibnamefont{Goan}},
  \bibinfo{author}{\bibfnamefont{T.~P.} \bibnamefont{Spiller}},
  \bibnamefont{and} \bibinfo{author}{\bibfnamefont{G.~J.}
  \bibnamefont{Milburn}}, \bibinfo{journal}{Phys. Rev. A}
  \textbf{\bibinfo{volume}{72}}, \bibinfo{pages}{062327}
  (\bibinfo{year}{2005}).

\bibitem[{\citenamefont{Hill and Ralph}(2008)}]{Hill2008}
\bibinfo{author}{\bibfnamefont{C.}~\bibnamefont{Hill}} \bibnamefont{and}
  \bibinfo{author}{\bibfnamefont{J.}~\bibnamefont{Ralph}},
  \bibinfo{journal}{Phys. Rev. A} \textbf{\bibinfo{volume}{77}},
  \bibinfo{pages}{014305} (\bibinfo{year}{2008}).

\bibitem[{\citenamefont{Martin et~al.}(2015)\citenamefont{Martin, Motzoi, Li,
  Sarovar, and Whaley}}]{HPFPRA}
\bibinfo{author}{\bibfnamefont{L.}~\bibnamefont{Martin}},
  \bibinfo{author}{\bibfnamefont{F.}~\bibnamefont{Motzoi}},
  \bibinfo{author}{\bibfnamefont{H.}~\bibnamefont{Li}},
  \bibinfo{author}{\bibfnamefont{M.}~\bibnamefont{Sarovar}}, \bibnamefont{and}
  \bibinfo{author}{\bibfnamefont{K.~B.} \bibnamefont{Whaley}},
  \bibinfo{journal}{Phys. Rev. A} \textbf{\bibinfo{volume}{92}},
  \bibinfo{pages}{062321} (\bibinfo{year}{2015}).

\bibitem[{\citenamefont{Jacobs}(2003)}]{Jacobs2003}
\bibinfo{author}{\bibfnamefont{K.}~\bibnamefont{Jacobs}},
  \bibinfo{journal}{Phys. Rev. A} \textbf{\bibinfo{volume}{67}}
  (\bibinfo{year}{2003}).

\bibitem[{\citenamefont{Wiseman and Ralph}(2006)}]{Wiseman2006reconsidering}
\bibinfo{author}{\bibfnamefont{H.~M.} \bibnamefont{Wiseman}} \bibnamefont{and}
  \bibinfo{author}{\bibfnamefont{J.}~\bibnamefont{Ralph}},
  \bibinfo{journal}{New J. Phys.} \textbf{\bibinfo{volume}{8}},
  \bibinfo{pages}{90} (\bibinfo{year}{2006}).

\bibitem[{\citenamefont{Wiseman and Bouten}(2008)}]{Wiseman2008}
\bibinfo{author}{\bibfnamefont{H.~M.} \bibnamefont{Wiseman}} \bibnamefont{and}
  \bibinfo{author}{\bibfnamefont{L.}~\bibnamefont{Bouten}},
  \bibinfo{journal}{Quantum Information Processing}
  \textbf{\bibinfo{volume}{7}}, \bibinfo{pages}{71} (\bibinfo{year}{2008}).

\bibitem[{\citenamefont{Teo et~al.}(2014)\citenamefont{Teo, Combes, and
  Wiseman}}]{Teo2014}
\bibinfo{author}{\bibfnamefont{C.}~\bibnamefont{Teo}},
  \bibinfo{author}{\bibfnamefont{J.}~\bibnamefont{Combes}}, \bibnamefont{and}
  \bibinfo{author}{\bibfnamefont{H.~M.} \bibnamefont{Wiseman}},
  \bibinfo{journal}{New J. Phys.} \textbf{\bibinfo{volume}{16}},
  \bibinfo{pages}{105010} (\bibinfo{year}{2014}).

\bibitem[{\citenamefont{Li et~al.}(2013)\citenamefont{Li, Shabani, Sarovar, and
  Whaley}}]{Li2013}
\bibinfo{author}{\bibfnamefont{H.}~\bibnamefont{Li}},
  \bibinfo{author}{\bibfnamefont{A.}~\bibnamefont{Shabani}},
  \bibinfo{author}{\bibfnamefont{M.}~\bibnamefont{Sarovar}}, \bibnamefont{and}
  \bibinfo{author}{\bibfnamefont{B.~K.} \bibnamefont{Whaley}},
  \bibinfo{journal}{Phys. Rev. A} \textbf{\bibinfo{volume}{87}},
  \bibinfo{pages}{032334} (\bibinfo{year}{2013}).

\bibitem[{\citenamefont{Combes and Jacobs}(2006)}]{Combes:2006hx}
\bibinfo{author}{\bibfnamefont{J.}~\bibnamefont{Combes}} \bibnamefont{and}
  \bibinfo{author}{\bibfnamefont{K.}~\bibnamefont{Jacobs}},
  \bibinfo{journal}{Phys. Rev. Lett.} \textbf{\bibinfo{volume}{96}},
  \bibinfo{pages}{010504} (\bibinfo{year}{2006}).

\bibitem[{\citenamefont{Shabani and Jacobs}(2008)}]{Shabani2008locally}
\bibinfo{author}{\bibfnamefont{A.}~\bibnamefont{Shabani}} \bibnamefont{and}
  \bibinfo{author}{\bibfnamefont{K.}~\bibnamefont{Jacobs}},
  \bibinfo{journal}{Phys. Rev. Lett.} \textbf{\bibinfo{volume}{101}},
  \bibinfo{pages}{230403} (\bibinfo{year}{2008}).

\bibitem[{\citenamefont{Combes et~al.}(2010)\citenamefont{Combes, Wiseman,
  Jacobs, and O’Connor}}]{Combes2010rapid}
\bibinfo{author}{\bibfnamefont{J.}~\bibnamefont{Combes}},
  \bibinfo{author}{\bibfnamefont{H.~M.} \bibnamefont{Wiseman}},
  \bibinfo{author}{\bibfnamefont{K.}~\bibnamefont{Jacobs}}, \bibnamefont{and}
  \bibinfo{author}{\bibfnamefont{A.~J.} \bibnamefont{O’Connor}},
  \bibinfo{journal}{Phys. Rev. A} \textbf{\bibinfo{volume}{82}},
  \bibinfo{pages}{022307} (\bibinfo{year}{2010}).

\bibitem[{\citenamefont{Sridharan et~al.}(2012)\citenamefont{Sridharan,
  Yanagisawa, and Combes}}]{Sridharan2012optimal}
\bibinfo{author}{\bibfnamefont{S.}~\bibnamefont{Sridharan}},
  \bibinfo{author}{\bibfnamefont{M.}~\bibnamefont{Yanagisawa}},
  \bibnamefont{and} \bibinfo{author}{\bibfnamefont{J.}~\bibnamefont{Combes}},
  \bibinfo{journal}{arXiv preprint arXiv:1211.5617}  (\bibinfo{year}{2012}).

\bibitem[{\citenamefont{Serafini and
  Mancini}(2010)}]{Serafini2010determination}
\bibinfo{author}{\bibfnamefont{A.}~\bibnamefont{Serafini}} \bibnamefont{and}
  \bibinfo{author}{\bibfnamefont{S.}~\bibnamefont{Mancini}},
  \bibinfo{journal}{Phys. Rev. Lett.} \textbf{\bibinfo{volume}{104}},
  \bibinfo{pages}{220501} (\bibinfo{year}{2010}).

\bibitem[{\citenamefont{Mancini and Wiseman}(2007)}]{Mancini2007optimal}
\bibinfo{author}{\bibfnamefont{S.}~\bibnamefont{Mancini}} \bibnamefont{and}
  \bibinfo{author}{\bibfnamefont{H.~M.} \bibnamefont{Wiseman}},
  \bibinfo{journal}{Phys. Rev. A} \textbf{\bibinfo{volume}{75}},
  \bibinfo{pages}{012330} (\bibinfo{year}{2007}).

\bibitem[{\citenamefont{Genoni et~al.}(2013)\citenamefont{Genoni, Mancini, and
  Serafini}}]{Genoni2013optimal}
\bibinfo{author}{\bibfnamefont{M.~G.} \bibnamefont{Genoni}},
  \bibinfo{author}{\bibfnamefont{S.}~\bibnamefont{Mancini}}, \bibnamefont{and}
  \bibinfo{author}{\bibfnamefont{A.}~\bibnamefont{Serafini}},
  \bibinfo{journal}{Phys. Rev. A} \textbf{\bibinfo{volume}{87}},
  \bibinfo{pages}{042333} (\bibinfo{year}{2013}).

\bibitem[{\citenamefont{Wootters}(1998)}]{Wootters1998}
\bibinfo{author}{\bibfnamefont{W.~K.} \bibnamefont{Wootters}},
  \bibinfo{journal}{Phys. Rev. Lett.} \textbf{\bibinfo{volume}{80}},
  \bibinfo{pages}{2245} (\bibinfo{year}{1998}), ISSN \bibinfo{issn}{0031-9007}.

\bibitem[{\citenamefont{Rist{\`e} et~al.}(2013)\citenamefont{Rist{\`e},
  Dukalski, Watson, de~Lange, Tiggelman, Blanter, Lehnert, Schouten, and
  DiCarlo}}]{Riste2013}
\bibinfo{author}{\bibfnamefont{D.}~\bibnamefont{Rist{\`e}}},
  \bibinfo{author}{\bibfnamefont{M.}~\bibnamefont{Dukalski}},
  \bibinfo{author}{\bibfnamefont{C.~A.} \bibnamefont{Watson}},
  \bibinfo{author}{\bibfnamefont{G.}~\bibnamefont{de~Lange}},
  \bibinfo{author}{\bibfnamefont{M.~J.} \bibnamefont{Tiggelman}},
  \bibinfo{author}{\bibfnamefont{Y.~M.} \bibnamefont{Blanter}},
  \bibinfo{author}{\bibfnamefont{K.~W.} \bibnamefont{Lehnert}},
  \bibinfo{author}{\bibfnamefont{R.~N.} \bibnamefont{Schouten}},
  \bibnamefont{and} \bibinfo{author}{\bibfnamefont{L.}~\bibnamefont{DiCarlo}},
  \bibinfo{journal}{Nature} \textbf{\bibinfo{volume}{502}},
  \bibinfo{pages}{350} (\bibinfo{year}{2013}).

\bibitem[{\citenamefont{Jozsa}(1994)}]{Jozsa1994fidelity}
\bibinfo{author}{\bibfnamefont{R.}~\bibnamefont{Jozsa}}, \bibinfo{journal}{J.
  Mod. Opt.} \textbf{\bibinfo{volume}{41}}, \bibinfo{pages}{2315}
  (\bibinfo{year}{1994}).

\bibitem[{\citenamefont{Wiseman and Milburn}(2009)}]{Wiseman2009}
\bibinfo{author}{\bibfnamefont{H.~M.} \bibnamefont{Wiseman}} \bibnamefont{and}
  \bibinfo{author}{\bibfnamefont{G.~J.} \bibnamefont{Milburn}},
  \emph{\bibinfo{title}{{Quantum measurement and control}}}
  (\bibinfo{publisher}{Cambridge University Press}, \bibinfo{year}{2009}).

\bibitem[{\citenamefont{Jacobs and Steck}(2006)}]{JacobsSteck2006}
\bibinfo{author}{\bibfnamefont{K.}~\bibnamefont{Jacobs}} \bibnamefont{and}
  \bibinfo{author}{\bibfnamefont{D.~A.} \bibnamefont{Steck}},
  \bibinfo{journal}{Contemp. Phys.} \textbf{\bibinfo{volume}{47}},
  \bibinfo{pages}{279} (\bibinfo{year}{2006}).

\bibitem[{\citenamefont{Jacobs}(2014)}]{jacobs2014quantum}
\bibinfo{author}{\bibfnamefont{K.}~\bibnamefont{Jacobs}},
  \emph{\bibinfo{title}{Quantum measurement theory and its applications}}
  (\bibinfo{publisher}{Cambridge University Press}, \bibinfo{year}{2014}).

\bibitem[{\citenamefont{{\O}ksendal}(2003)}]{Oksendal2003}
\bibinfo{author}{\bibfnamefont{B.}~\bibnamefont{{\O}ksendal}},
  \emph{\bibinfo{title}{{Stochastic Differential Equations}}}
  (\bibinfo{publisher}{Spinger-Verlag}, \bibinfo{year}{2003}).

\bibitem[{\citenamefont{Nielsen and Chuang}(2010)}]{Nielsen2010}
\bibinfo{author}{\bibfnamefont{M.~A.} \bibnamefont{Nielsen}} \bibnamefont{and}
  \bibinfo{author}{\bibfnamefont{I.~L.} \bibnamefont{Chuang}},
  \emph{\bibinfo{title}{{Quantum computation and quantum information}}}
  (\bibinfo{publisher}{Cambridge University Press}, \bibinfo{year}{2010}).

\bibitem[{\citenamefont{Jacobs and Knight}(1998)}]{Jacobs1998linear}
\bibinfo{author}{\bibfnamefont{K.}~\bibnamefont{Jacobs}} \bibnamefont{and}
  \bibinfo{author}{\bibfnamefont{P.}~\bibnamefont{Knight}},
  \bibinfo{journal}{Phys. Rev. A} \textbf{\bibinfo{volume}{57}},
  \bibinfo{pages}{2301} (\bibinfo{year}{1998}).

\bibitem[{\citenamefont{Jacobs and Shabani}(2008)}]{Shabani2008}
\bibinfo{author}{\bibfnamefont{K.}~\bibnamefont{Jacobs}} \bibnamefont{and}
  \bibinfo{author}{\bibfnamefont{A.}~\bibnamefont{Shabani}},
  \bibinfo{journal}{Contemp. Phys.} \textbf{\bibinfo{volume}{49}},
  \bibinfo{pages}{435} (\bibinfo{year}{2008}).

\bibitem[{\citenamefont{Verstraete and
  Verschelde}(2002)}]{Verstraete2002fidelity}
\bibinfo{author}{\bibfnamefont{F.}~\bibnamefont{Verstraete}} \bibnamefont{and}
  \bibinfo{author}{\bibfnamefont{H.}~\bibnamefont{Verschelde}},
  \bibinfo{journal}{Phys. Rev. A} \textbf{\bibinfo{volume}{66}},
  \bibinfo{pages}{022307} (\bibinfo{year}{2002}).

\bibitem[{\citenamefont{Clerk et~al.}(2010)\citenamefont{Clerk, Girvin,
  Marquardt, and Schoelkopf}}]{Clerk:2010dh}
\bibinfo{author}{\bibfnamefont{A.~A.} \bibnamefont{Clerk}},
  \bibinfo{author}{\bibfnamefont{S.~M.} \bibnamefont{Girvin}},
  \bibinfo{author}{\bibfnamefont{F.}~\bibnamefont{Marquardt}},
  \bibnamefont{and} \bibinfo{author}{\bibfnamefont{R.~J.}
  \bibnamefont{Schoelkopf}}, \bibinfo{journal}{Rev. Mod. Phys.}
  \textbf{\bibinfo{volume}{82}}, \bibinfo{pages}{1155} (\bibinfo{year}{2010}).

\bibitem[{\citenamefont{Shankar et~al.}(2013)\citenamefont{Shankar, Hatridge,
  Leghtas, Sliwa, Narla, Vool, Girvin, Frunzio, Mirrahimi, and
  Devoret}}]{Shankar2013autonomously}
\bibinfo{author}{\bibfnamefont{S.}~\bibnamefont{Shankar}},
  \bibinfo{author}{\bibfnamefont{M.}~\bibnamefont{Hatridge}},
  \bibinfo{author}{\bibfnamefont{Z.}~\bibnamefont{Leghtas}},
  \bibinfo{author}{\bibfnamefont{K.}~\bibnamefont{Sliwa}},
  \bibinfo{author}{\bibfnamefont{A.}~\bibnamefont{Narla}},
  \bibinfo{author}{\bibfnamefont{U.}~\bibnamefont{Vool}},
  \bibinfo{author}{\bibfnamefont{S.~M.} \bibnamefont{Girvin}},
  \bibinfo{author}{\bibfnamefont{L.}~\bibnamefont{Frunzio}},
  \bibinfo{author}{\bibfnamefont{M.}~\bibnamefont{Mirrahimi}},
  \bibnamefont{and} \bibinfo{author}{\bibfnamefont{M.~H.}
  \bibnamefont{Devoret}}, \bibinfo{journal}{Nature}
  \textbf{\bibinfo{volume}{504}}, \bibinfo{pages}{419} (\bibinfo{year}{2013}).

\bibitem[{\citenamefont{Kimchi-Schwartz
  et~al.}(2016)\citenamefont{Kimchi-Schwartz, Martin, Flurin, Aron, Kulkarni,
  Tureci, and Siddiqi}}]{Schwartz2015}
\bibinfo{author}{\bibfnamefont{M.~E.} \bibnamefont{Kimchi-Schwartz}},
  \bibinfo{author}{\bibfnamefont{L.}~\bibnamefont{Martin}},
  \bibinfo{author}{\bibfnamefont{E.}~\bibnamefont{Flurin}},
  \bibinfo{author}{\bibfnamefont{C.}~\bibnamefont{Aron}},
  \bibinfo{author}{\bibfnamefont{M.}~\bibnamefont{Kulkarni}},
  \bibinfo{author}{\bibfnamefont{H.~E.} \bibnamefont{Tureci}},
  \bibnamefont{and} \bibinfo{author}{\bibfnamefont{I.}~\bibnamefont{Siddiqi}},
  \bibinfo{journal}{Phys. Rev. Lett.} \textbf{\bibinfo{volume}{116}},
  \bibinfo{pages}{240503} (\bibinfo{year}{2016}).

\bibitem[{\citenamefont{Ticozzi and Viola}(2009)}]{ticozzi2009analysis}
\bibinfo{author}{\bibfnamefont{F.}~\bibnamefont{Ticozzi}} \bibnamefont{and}
  \bibinfo{author}{\bibfnamefont{L.}~\bibnamefont{Viola}},
  \bibinfo{journal}{Automatica} \textbf{\bibinfo{volume}{45}},
  \bibinfo{pages}{2002} (\bibinfo{year}{2009}).

\bibitem[{\citenamefont{Ticozzi and Viola}(2012)}]{ticozzi2012stabilizing}
\bibinfo{author}{\bibfnamefont{F.}~\bibnamefont{Ticozzi}} \bibnamefont{and}
  \bibinfo{author}{\bibfnamefont{L.}~\bibnamefont{Viola}},
  \bibinfo{journal}{Philosophical Transactions of the Royal Society of London
  A: Mathematical, Physical and Engineering Sciences}
  \textbf{\bibinfo{volume}{370}}, \bibinfo{pages}{5259} (\bibinfo{year}{2012}).

\bibitem[{\citenamefont{Johnson et~al.}(2015)\citenamefont{Johnson, Ticozzi,
  and Viola}}]{JohnsonViola2015}
\bibinfo{author}{\bibfnamefont{P.~D.} \bibnamefont{Johnson}},
  \bibinfo{author}{\bibfnamefont{F.}~\bibnamefont{Ticozzi}}, \bibnamefont{and}
  \bibinfo{author}{\bibfnamefont{L.}~\bibnamefont{Viola}},
  \bibinfo{journal}{arXiv preprint arXiv:1506.07756}  (\bibinfo{year}{2015}).

\bibitem[{\citenamefont{Liu et~al.}(2016)\citenamefont{Liu, Shankar, Ofek,
  Hatridge, Narla, Sliwa, Frunzio, Schoelkopf, and Devoret}}]{liu2016comparing}
\bibinfo{author}{\bibfnamefont{Y.}~\bibnamefont{Liu}},
  \bibinfo{author}{\bibfnamefont{S.}~\bibnamefont{Shankar}},
  \bibinfo{author}{\bibfnamefont{N.}~\bibnamefont{Ofek}},
  \bibinfo{author}{\bibfnamefont{M.}~\bibnamefont{Hatridge}},
  \bibinfo{author}{\bibfnamefont{A.}~\bibnamefont{Narla}},
  \bibinfo{author}{\bibfnamefont{K.}~\bibnamefont{Sliwa}},
  \bibinfo{author}{\bibfnamefont{L.}~\bibnamefont{Frunzio}},
  \bibinfo{author}{\bibfnamefont{R.~J.} \bibnamefont{Schoelkopf}},
  \bibnamefont{and} \bibinfo{author}{\bibfnamefont{M.~H.}
  \bibnamefont{Devoret}}, \bibinfo{journal}{Phys. Rev. X}
  \textbf{\bibinfo{volume}{6}}, \bibinfo{pages}{011022} (\bibinfo{year}{2016}).

\bibitem[{\citenamefont{Bennett et~al.}(1996)\citenamefont{Bennett, Brassard,
  Popescu, Schumacher, Smolin, and Wootters}}]{Wootters1997}
\bibinfo{author}{\bibfnamefont{C.~H.} \bibnamefont{Bennett}},
  \bibinfo{author}{\bibfnamefont{G.}~\bibnamefont{Brassard}},
  \bibinfo{author}{\bibfnamefont{S.}~\bibnamefont{Popescu}},
  \bibinfo{author}{\bibfnamefont{B.}~\bibnamefont{Schumacher}},
  \bibinfo{author}{\bibfnamefont{J.~A.} \bibnamefont{Smolin}},
  \bibnamefont{and} \bibinfo{author}{\bibfnamefont{W.~K.}
  \bibnamefont{Wootters}}, \bibinfo{journal}{Phys. Rev. Lett.}
  \textbf{\bibinfo{volume}{76}} (\bibinfo{year}{1996}).

\end{thebibliography}

\clearpage
\end{document}